\documentclass[a4paper,10pt,final,twocolumn]{IEEEtran}
\usepackage{graphics, graphicx}
\usepackage{amssymb,amsmath}
\usepackage{amsmath}
\usepackage[rm]{subfigure}    
\usepackage{threeparttable}   
\usepackage[dvips]{color}
\usepackage{url}   
\usepackage[latin1]{inputenc} 
\usepackage[T1]{fontenc}
\usepackage{algorithm}               % Specifies the index style.
\usepackage{cite}
\usepackage{multirow}
\usepackage{lineno}
\usepackage[normalem]{ulem}
\usepackage{soul}  % use \st{works}, where works will be marked with a medium line across the works
\usepackage{float}
\usepackage{flafter}
\usepackage{stfloats}
\usepackage{url}

\usepackage[outerbars,color]{changebar}
\ifx\pdfoutput\undefined
\else\ifnum\pdfoutput>0
  \usepackage{pdfcolmk}
\fi\fi
\cbcolor{blue}
\usepackage{fancyhdr}
\usepackage{amsmath}
\DeclareMathOperator*{\argmax}{arg\,max}
\DeclareMathOperator*{\argmin}{arg\,min}

\usepackage{algorithm}
\usepackage{indentfirst}
\usepackage{setspace}

\usepackage{tikz}
\definecolor{gold}{rgb}{0.85,.66,0}

\newcommand{\bm}[1]{\mbox{\boldmath{$#1$}}}
\newcommand{\eqdef}{\ensuremath{\mathrel{\stackrel{\mathrm{def}}{=}}}}

\renewcommand{\IEEEQED}{\IEEEQEDopen}
\newtheorem{lemma}{Lemma}

\begin{document}
%==============================================================
\title{Energy Efficient OFDMA Networks Maintaining Statistical QoS Guarantees for Delay-Sensitive Traffic }
%==============================================================
\author{Taufik~Abrão, {\em Senior Member,~IEEE}, Lucas~D.~H.~Sampaio, Shaoshi~Yang, Kent~Tsz~Kan~Cheung, 
Paul~Jean~E.~Jeszensky, {\em Senior Member,~IEEE} and Lajos~Hanzo, {\em Fellow,~IEEE}
\thanks{The financial support of the National Council for Scientific and Technological Development (CNPq) of Brazil under Grants 202340/2011-2 and 303426/2009-8, of the Londrina State University (UEL) and the Paraná State Government, of the EPSRC projects EP/N004558/1 and
EP/L018659/1, as well as of the European Research Council's Advanced Fellow Grant under the Beam-Me-Up project is gratefully acknowledged.}
\thanks{T. Abrão is with the Electrical Engineering Department, Londrina State University, Paraná, Brazil, and was also with the School of Electronics and Computer Science, University of Southampton, SO17 1BJ Southampton, U.K. (e-mail: taufik@uel.br).}
\thanks{L. D. H. Sampaio and P. J. E. Jeszensky are with the Polytechnic School of the University of São Paulo (EPUSP), Brazil (e-mail:  \{ldhsampaio, pjej\}@usp.br).}
\thanks{S. Yang, K. T. K. Cheung and L. Hanzo are with the School of Electronics and Computer Science, University of Southampton, SO17 1BJ Southampton, U.K. (e-mail: \{sy7g09, ktkc106, lh\}@ecs.soton.ac.uk).}
}

\markboth{Accepted to appear on IEEE Access, Jan. 2016}%
{Shell \MakeLowercase{\textit{et al.}}: Bare Demo of IEEEtran.cls
for Journals}

\maketitle
%\linespread{2.1}
%\linespread{1.7}
\textheight = 255mm 
% \textwidth=200mm 
% \evensidemargin=-2cm

%\colr{\bf red: Warning! It needs action/correction/attention of ALL!}
\vspace{-1cm}
\begin{abstract}
An energy-efficient design is proposed under specific statistical quality-of-service (QoS) guarantees for delay-sensitive traffic in the downlink orthogonal frequency-division multiple-access (OFDMA) networks. This design is based on Wu's \textit{effective capacity} (EC) concept\cite{Wu03}, which characterizes the maximum throughput of a system subject to statistical delay-QoS requirements at the data-link layer. In the particular context considered, our main contributions consist of quantifying the \textit{effective energy-efficiency} (EEE)-versus-EC tradeoff and characterizing  the delay-sensitive traffic as a function of the QoS-exponent $\theta$, which expresses the exponential decay rate of the delay-QoS violation probabilities. Upon exploiting the properties of fractional programming, the originally quasi-concave EEE optimization problem having a fractional form is transformed into a subtractive optimization problem by applying Dinkelbach's method. As a result, an iterative inner-outer loop based resource allocation algorithm is conceived for efficiently solving the transformed EEE optimization problem. Our simulation results demonstrate that the proposed scheme converges within a few Dinkelbach algorithm's iterations to the desired solution accuracy. Furthermore, the impact of the circuitry power, of the QoS-exponent and of the power amplifier inefficiency is characterized numerically. These results reveal that the optimally allocated power maximizing the EEE decays exponentially with respect to both the circuitry power and the QoS-exponent, whilst decaying linearly with respect to the power amplifier inefficiency.
\end{abstract}


%========================
%%%%%%%%%%%%%%%%%%%%%%%%%
\section{Introduction}
%%%%%%%%%%%%%%%%%%%%%%%%%
%========================
\subsection{Motivations}

It is predicted that a formidable 1000-fold mobile data traffic growth and a near-zero latency have to be met by the forthcoming fifth generation (5G) mobile communication systems\cite{Qualcomm_1000x, Huawei_5G_whitepaper}, which are expected to support bandwidth-thirsty delay-sensitive multimedia services, such as ultra high-definition (UHD) video streaming\cite{Yang_2015:LS_MIMO_HEVC}. Meanwhile, the economical, environmental and societal pressures require a significant reduction of the carbon-footprint of the ubiquitous information and communication technologies (ICT), which will be responsible for 4 - 6\% of the annual global greenhouse gas emissions by 2020, unless the energy-consumption-per-bit is sharply reduced\cite{Fujitsu_ICT_energy}. Conventional designs of wireless communication networks have been  dominated by improving the attainable spectral efficiency (SE), which was achieved by degrading the 5G design objectives concerning the energy-efficiency (EE) and delay. Therefore, an important research challenge for sustainable future wireless communication systems has been how to achieve significantly higher throughput (bits/second), while simultaneously improving the energy-efficiency (EE) and the delay.

According to the Shannon-Hartley theorem\cite{Shannon_1948:math_theory_comm}, in a point-to-point signal link having a given bandwidth $W$ and additive white Gaussian noise (AWGN) power spectral density (PSD) $N_0$, the maximum achievable transmission rate $R$ [bits/second] of this link is logarithmically proportional to the transmit power $P$: 
\begin{equation}
 R = W\log_2\left(1 + \frac{P}{N_0 W}\right). 
\end{equation}
Therefore, the relationship between the SE  $\eta_{SE} = \frac{R}{W}$ [bits/second/Hz] and EE $\eta_{EE} = \frac{R}{P}$ [bits/second/Watt or bits/Joule] can be expressed as\footnote{The definition of EE has several variants. By analogy with the definition of SE, the EE defined here can also be interpreted as power efficiency (PE), which is in fact used interchangeably with EE in the open literature and we follow this convention in this paper unless stated otherwise.} 
\begin{equation}
 \eta_{EE} = \frac{\eta_{SE}}{(2^{\eta_{SE}} -1)N_0 }. 
\end{equation}
It is plausible that when $\eta_{SE}$ approaches zero, $\eta_{EE}$ converges to a constant $\frac{1}{N_0\ln 2}$; while if $\eta_{SE}$ tends to infinity, $\eta_{EE}$ approaches zero\cite{Chen_2011:green_comm_tradeoff}. 
 As a result, in general the SE and EE of a communication system conflict with each other. 

In order to achieve a desirable EE-SE tradeoff (EST), radio resources such as the available transmit power and bandwidth (e.g. the subcarriers in orthogonal frequency-division multiple-access (OFDMA), which has been used in LTE-family of wireless standards), have to be appropriately allocated to different users.

\subsection{Related Works}
The SE-maximization problem has been studied in various contexts during the last few decades. By contrast, the EE-maximization became a hot topic in the resource allocation (RA) of wireless communication systems only recently. For instance, in \cite{Xiong11}, a general EST framework was proposed for the downlink OFDMA networks, where the overall EE, SE and per-user rate constraints were jointly considered, while a tight upper bound and lower bound on the optimal EST relationship were obtained  based on Lagrangian dual decomposition. Additionally, it was demonstrated under this framework that the EE is a strictly quasi-concave function of the SE\cite{Xiong11}. Furthermore, energy-efficient RA in both the downlink and uplink of cellular OFDMA networks has been studied in \cite{Xiong_2012_EE_maximization_OFDMA}. Explicitly, for the downlink transmission the weighted EE was maximized, while for the uplink it was the minimum individual EE that was maximized, both under certain prescribed per-user rate requirements. As a further advance, a series of optimization problems concerning both the SE and the spectral-normalized EE [bits/Joule/Hz] maximization in the context of multi-relay aided OFDMA networks subject to a maximum total network transmit power budget were studied in\cite{Kent_2013:GC_EEM, Kent_2013:EEM_fractional_programming, Kent_2014:EEM_MIMO_OFDMA, Kent_TSP_2015:EE_IA_multicell}. To elaborate a little further, \cite{Kent_2013:GC_EEM, Kent_2013:EEM_fractional_programming} considered the scenario where each network entity has only a single antenna, and the classic Dinkelbach's method was invoked for solving the resultant fractional programming problem. By contrast, \cite{Kent_2014:EEM_MIMO_OFDMA} considered the more complex and generalized context where each network entity is equipped with multiple antennas, and the low-complexity Charnes-Cooper transformation method was employed for solving the resultant fractional programming problem. Furthermore, the EE optimization problem for the most complicated multi-cell multi-antenna multi-relay OFDMA networks was studied in\cite{Kent_TSP_2015:EE_IA_multicell}. To achieve the optimum SE and/or EE, the emerging interference alignment (IA) technique was adopted for managing the multi-cell co-channel interference, which represents the first work having studied the EE of IA techniques. Another interesting contribution was provided in \cite{Wenpeng_2015:flexible_EE_SE_optimization}, where a multi-cell OFDMA network was considered, and a novel EST metric capable of simultaneously capturing both the EST relationship and the individual cells' preferences for the EE or SE performance, was introduced as the utility function for each base station (BS).    

However, the system's delay, which is a vitally important quality-of-service (QoS) metric for delay-sensitive multimedia applications in 5G communications, was not considered 
in\cite{Xiong11, Xiong_2012_EE_maximization_OFDMA, Kent_2013:GC_EEM, Kent_2013:EEM_fractional_programming, Kent_2014:EEM_MIMO_OFDMA, Wenpeng_2015:flexible_EE_SE_optimization, Kent_TSP_2015:EE_IA_multicell}. Since the achievable data rate varies as a function of the fading channel's quality, satisfying deterministic delay-QoS constraints is quite challenging, even impossible in some cases. As a result, satisfying statistical delay-QoS specifications for transmission over wireless channels becomes relevant, when the delay of certain services must be lower than a specific threshold for at least a certain percentage of time\cite{Chang94, Chang95}. 

Most of existing delay-QoS related contributions did not consider the system's EE\cite{Tang07, Tang07b, Tang08, Musavian10, Du11, Du2011b}. For example, in \cite{Tang07} the data-link layer's delay-QoS performance was characterized using a cross-layer model relying on the \textit{effective capacity (EC)} concept\cite{Wu03}, which has been recognized as a critically important metric for the statistical delay-QoS guarantees in wireless mobile networks. Based on this cross-layer model, a pair of adaptive RA schemes aiming for achieving the maximum EC over single-hop fading wireless links were proposed in\cite{Tang07b, Tang08}. Additionally, the authors of \cite{Musavian10} investigated the EC of a cognitive radio relay network, when the secondary user transmission is subject to satisfying spectrum-sharing restrictions imposed by a primary user. The authors of \cite{Du11} proposed a delay-QoS-driven power allocation scheme for two-hop wireless relay links, while a delay-QoS-driven BS selection algorithm was proposed in \cite{Du2011b} for satisfying multiple downlink users' delay-bound violation probabilities. 

Nonetheless, there are a few seminal contributions related to the EE of delay-constrained systems. For example, in \cite{Zhang11} the overall transmit power of vehicle-to-roadside infrastructure communication networks was minimized by jointly assigning power and subcarriers under delay-aware QoS requirements. More specifically, the authors of \cite{Zhang11} developed a cross-layer framework where orthogonal frequency-division multiplexing (OFDM), which may be regarded as a special case of OFDMA, was employed at the physical layer, while the power- and the subcarrier-assignment policy operates at the data-link layer. Additionally, in \cite{Loodaricheh2014} an energy-efficient RA scheme was proposed for multiuser cooperation aided OFDMA networks under a specific rate-QoS provision. To elaborate a little further, in \cite{Loodaricheh2014} a joint power allocation, subcarrier allocation as well as mobile-relay selection algorithm was developed, aiming for maximizing the system's overall EE by taking into account different rate-QoS requirements. The authors of \cite{Qiang_Ni_2015} indeed investigated the \textit{effective energy efficiency (EEE)} maximization under the EC-based statistical delay-QoS constraint. However, they considered a simple point-to-point communication system, where only power allocation is involved\cite{Qiang_Ni_2015}.

\subsection{Contributions of This Paper}
Against the above background, in this paper we propose an energy-efficient RA strategy under a specific statistical delay-QoS provision for delay-sensitive applications in the downlink of OFDMA cellular networks. Furthermore, the impact of the circuitry power, of the QoS-exponent and of the power amplifier inefficiency is characterized numerically. These results reveal that the optimally allocated power maximizing the EEE decays exponentially both with the circuitry power and with the QoS-exponent, whilst decaying linearly with respect to the power amplifier inefficiency. The main contributions of this paper are significantly different from those of \cite{Loodaricheh2014}, although it is probably the most closely related work to ours.  
\begin{itemize}
\item We consider a non-cooperative OFDMA network, while the RA in \cite{Loodaricheh2014} was carried out by considering a user-cooperation aided OFDMA network relying on time-division duplex (TDD). 
\item In the cross-layer optimization problem considered, only channel statistics are needed for obtaining both power- and subcarrier-allocation solutions, while the instantaneous channel state information (CSI) was required by the RA scheme of \cite{Loodaricheh2014}. As a result, our approach significantly simplifies the RA strategy to be used in the OFDMA networks that are capable of supporting delay-sensitive traffic.
\item Our work invokes the EC concept instead of Shannon's channel capacity. As a result, we investigate the tradeoff between the EEE and the EC. By contrast, in most existing literature, such as \cite{Loodaricheh2014}, the tradeoff between the traditional EE and SE was studied.    
\item In the particular optimization problem solved in this paper, the maximum delay bound and the probability of delay-QoS violation are characterized jointly with the aid of the statistical QoS-exponent $\theta$. Furthermore, the minimum EC constraint is also investigated and incorporated in our optimization problem (not as a delay constraint though).  By contrast, statistical delay-QoS concept was not considered in \cite{Loodaricheh2014}, where the delay tolerance was in fact implicitly mapped to a traditional minimum-rate requirement.
\end{itemize}

The remainder of this paper is organized as follows. The preliminaries and an OFDMA power consumption model are introduced in Section \ref{sec:prelim}. In Section \ref{sec:probl_formul}, the EEE optimization problem is formulated. The solution approach combining Dinkelbach's method and Lagrangian dual decomposition is presented in Section \ref{sec:solution}. Our numerical simulation results are provided in Section \ref{sec:numerical}, which demonstrated the efficacy of the proposed algorithm. Finally, Section \ref{sec:conclusions} concludes the paper.

%========================
%%%%%%%%%%%%%%%%%%%%%%%%%
\section{Preliminaries}\label{sec:prelim}
%%%%%%%%%%%%%%%%%%%%%%%%%
%========================
%---------------------------------
In this section, the data-link layer queueing model, the major concepts regarding the statistical delay-QoS guarantee, and the power consumption model invoked are briefly revisited for making the paper self-contained. 
\subsection{Queueing, Effective Bandwidth (EB) and Effective Capacity (EC)}
%---------------------------------
There are two important concepts associated with the data-link layer's delay-bound violation probability, namely the EB\cite{Chang95b} and the EC\cite{Wu03}. Both of them rely on the queueing (first-in first-out buffering) model, which is employed for matching the \textit{source traffic arrival process} and the \textit{network service process}. As a benefit of the buffer, the queue prevents the loss of packets that could take place when the source rate is higher than the service rate, which is achieved at the expense of an increased delay.     

\subsubsection{Queuing-induced Delay} Assuming stationary arrival and service processes, at a given time instant $t$, the parameter $\theta$, which is the so-called ``QoS-exponent'' representing the decay rate of the tail distribution of the queue length $Q(t)$, satisfies\cite{Chang94, Chang95}:
\begin{equation}
	\lim_{q\rightarrow \infty} \frac{\ln \, {\rm Pr}[Q(t) \geq q] }{q} = -\theta.
\end{equation}
In other words, the probability of the queue length exceeding a certain threshold $q$ decays exponentially as the threshold $q$ increases. As a consequence, given a sufficiently large {\it maximum tolerable stationary queue length} $q_{\max}$, the following approximation is valid for the buffer-overflow probability \cite{Chang94}:
\begin{equation}\label{eq:large_queue_length}
{\rm Pr}[Q(t) \geq q_{\max}] \approx e^{-q_{\max}\theta }. 
\end{equation}
By contrast, for a small $q_{\max}$, the following approximation was shown to be more accurate\cite{Wu03}:
\begin{equation}\label{eq:queue_length_small}
{\rm Pr}[Q(t) \geq q_{\max}] \approx \alpha e^{-q_{\max}\theta },
\end{equation}
where $\alpha = {\rm Pr}[Q(t) \geq 0] $ denotes the probability that the buffer is not empty, which is approximated by the ratio of the average arrival rate over the average service rate\cite{Chang95}.  

Similarly, when the QoS metric of interest is \textit{delay}, with $D(t)$ denoting the delay experienced by a source packet arriving at time instant $t$ with respect to the buffer, and upon assuming a maximum tolerable delay of $d_{\max}$ [second], the following approximation holds:
\begin{equation}\label{eq:dmax}
{\rm Pr}[D(t) \geq d_{\max}] \approx \alpha e^{-\theta \delta d_{\max}} \leq \varepsilon, 
\end{equation}
where $\delta$  is the fixed rate [bits/second] jointly determined by the arrival and service processes relying on a relationship between EB and EC, as detailed later. Explicitly, \eqref{eq:dmax} indicates that the delay-bound violation probability must not be higher than  $\varepsilon$. 
To elaborate a little further, a smaller $\theta$ implies a slower rate of decay, which indicates that the system can only provide a looser delay-QoS guarantee. By contrast, a larger $\theta$ results in a faster rate of decay, which implies that a more stringent delay-QoS requirement can be supported. In particular, when $\theta\rightarrow \infty$, the system can tolerate an arbitrarily long delay. On the other hand, when $\theta\rightarrow 0$, the system cannot tolerate ``any'' delay, which corresponds to an extremely stringent delay-bound. The statistical delay-QoS constraint of \eqref{eq:dmax} may also be interpreted as the packet loss rate (PLR) requirement \cite{Ahn10}, because once the buffer is full and the delay is in excess of its maximum, the packets have to be dropped. Based on this relationship, from \eqref{eq:dmax}, the QoS-exponent for a certain user can be bounded as:
\begin{equation}\label{eq:theta_bound}
\theta \geq \frac{-\ln \varepsilon} {\delta d_{\max}}  \qquad\qquad \left[\frac{1}{\text{bits}} \right].
\end{equation}
When the \emph{delay bound} $d_{\max}$ is the main QoS metric of interest, we can further define the \emph{delay-QoS-exponent} as $\theta_\textsc{d} = \theta \delta = - \frac{\ln \varepsilon}{d_{\max}}$.

\subsubsection{Concepts of EB and EC} The QoS-exponent $\theta > 0$ or the delay-QoS-exponent $\theta_\textsc{d}$ is of paramount importance in terms of characterizing the statistical delay-QoS guarantees, since they both characterize the exponential decay rate of the delay-QoS violation probabilities.   

The stochastic behavior of a source traffic arrival process can be modeled asymptotically by its EB function $\mathcal{B}_{\rm e} (\theta)$. More specifically, let us consider an arrival process $\{A(t), t \ge 0\}$, where $A(t)$ represents the amount of source data [bits] arriving over the time interval $[0, t)$. Let us assume that the Gärtner-Ellis limit of the arrival process $A(t)$, which is defined as the asymptotic log-moment generating function of $A(t)$:
\begin{eqnarray}
	\Lambda_B(\theta)  &\eqdef& \lim_{t \rightarrow \infty} \quad \frac{1}{t} \ln \mathbb{E}\left(e^{\theta A(t)}\right) ,
\end{eqnarray}
does exist for all $\theta \geq0$ and that $\Lambda_B(\theta)$ is differentiable. Then, the EB function of $A(t)$ is defined as \cite{Wu03, Chang95b}:
\begin{equation}\label{eq:EB}
\mathcal{B}_{\rm e}(\theta)=\frac{\Lambda_B(\theta) }{\theta},\,\, \forall \theta \geq0. 
\end{equation}

Analogously to the arrival process $A(t)$, let the sequence $\{R[i], \, i= 1, 2,\ldots\}$ represent a discrete-time stationary and ergodic stochastic service process and $\mathcal{R}[t] \eqdef \sum_i^t R[i]$ be the partial sum of this discrete-time stochastic service process\footnote{Note that the service provided by the channel can be calculated, in a continuous-time, as $\mathcal{R}(t) = \int_0^t \texttt{r}(\tau)d\tau$, where $\texttt{r}(t)$ is the instantaneous capacity of the channel at time $t$. Furthermore, we emphasize that in the RA scheme proposed, it is not necessary to explicitly calculate $\texttt{r}(t)$, hence no instantaneous CSI knowledge is required by our RA scheme. Instead, since the optimization problem formulated is based on statistical expectation of the delay tolerance and its violation probability, only the channel statistics have to be known at the base station. Similarly, we do not have to explicitly calculate \eqref{eq:rate_k_n} and \eqref{eq:p_k_n}.}, which represents the data [bits] communicated over the time sequence of $i=1,2,\ldots, t$. Furthermore, we assume that the asymptotic log-moment generating function of the service process $\mathcal{R}[t]$, which is defined as
\begin{eqnarray}
	\Lambda_C(\theta)  &=& \lim_{t \rightarrow \infty} \quad \frac{1}{t} \ln \mathbb{E}\left(e^{\theta \mathcal{R}[t]}\right),
\end{eqnarray}
does exist for all $\theta \geq0$ and that it is differentiable for all $\theta \in \mathbb{R}$ \cite{Chang94}, where $\mathbb{E}(\cdot)$ is the expectation operator with respect to $\mathcal{R}[t]$. Additionally, we assume that $\Lambda_C(\theta)$ is a convex function.  Then, the EC function of the service process $\mathcal{R}[t]$ under a given statistical delay-QoS requirement specified by the exponent $\theta>0$ is defined as \cite{Wu03}:
\begin{eqnarray}\label{eq:efcap}
	\mathcal{C}_{\rm e}(\theta) &\eqdef& -\frac{\Lambda_C(-\theta)}{\theta} 
	=  -\lim_{t \rightarrow \infty} \, \frac{1}{t\theta} \ln \mathbb{E}\left(e^{-\theta \mathcal{R}[t]}\right).
\end{eqnarray}

It should be noted that when the sequence $\{R[i], \, i=1, 2,\ldots\}$ associated with the service process $\mathcal{R}[t]$ is a statistically uncorrelated process\footnote{For instance, a communication process taking place over block-fading channels. In this case, $i = 1, 2, \cdots$ represents the indices of the fading blocks.}, the EC expression of \eqref{eq:efcap} may be simplified as:
\begin{equation}\label{eq:efcap_uncor}
	\mathcal{C}_{\rm e}(\theta) = -\frac{1}{\theta} \ln \mathbb{E}\left(e^{-\theta R[i]}\right),
\end{equation}
It is important to note that the EC in \eqref{eq:efcap_uncor} is a monotonically decreasing function of $\theta$ \cite{Tang07}, \cite{Zhang11}. 

\textit{Remark:} The QoS of a user may be uniquely and unambiguously specified by the statistical QoS-triplet $(\delta, d_{\max}, \alpha)$, and the EB may be interpreted as the \emph{minimum constant service rate} required by a given arrival process for which the QoS-exponent $\theta$ is fulfilled \cite{Tang07b}. Hence, the EC may be regarded as the dual concept of the EB.  Since its inception, the EC has become an important data-link layer metric that provides unique insights into the entire network's performance in the presence of statistical delay-QoS limitations. 

The classic \textit{large deviations theory} was employed for the formulation of the EC, which incorporates the statistical delay-QoS constraints by capturing the decay rate of the buffer occupancy probability for large queue lengths. Since the average arrival rate is equal to the average departure/service rate when the queue is in its steady-state\footnote{This condition is satisfied when large $q_{\max}$ and $d_{\max}$ are considered, and it also implies that $\alpha$  in \eqref{eq:queue_length_small} is \textit{almost surely} equal to one.}, the EC can be physically interpreted as the maximum throughput of a system whose queue is in its steady-state\cite{Qiao09}, subject to the constraints imposed on the queue length/buffer-overflow probability of \eqref{eq:large_queue_length} or similarly on the delay-bound violation probability of \eqref{eq:dmax}, where $\alpha$ is \textit{almost surely} equal to one. Viewed from a different perspective, the EC may also be interpreted as the maximum attainable service-rate as a function of the QoS-exponent $\theta\geq 0$, or as the maximum constant arrival rate that a given service process is capable of coping with, whilst  guaranteeing a statistical delay-QoS requirement specified by $\theta\geq 0$. 

The relationship between $\mathcal{C}_{\rm e}(\theta)$ of \eqref{eq:efcap_uncor} and $\mathcal{B}_{\rm e}(\theta)$ of \eqref{eq:EB} has been extensively characterized in\cite{Wu03, Tang07, Tang07b, Zhang11}. More specifically, as demonstrated in \cite{Tang07}, the EB and EC exhibit opposite trends when the QoS-exponent $\theta$ varies, i.e. $\mathcal{C}_{\rm e}(\theta)$ decreases with $\theta$ while $\mathcal{B}_{\rm e}(\theta)$ increases with $\theta$. As a result, there exists a crossing-point between the EB and EC curves, which implies that the achievable rate and the QoS-exponent solution pair ($\delta, \theta^*$) may be obtained by satisfying $\mathcal{C}_{\rm e}(\theta^*)=\mathcal{B}_{\rm e}(\theta^*) = \delta$.

%and is shown in Fig.~\ref{fig:CeBe} for clarity, where $\delta$ and $\theta^*$ is obtained such that $\Lambda_B(\theta^*)  = \mathcal{C}_{\rm e}(\theta^*) = \delta$.  
% \vspace{-5mm}
% \begin{figure}[t]
% \centering
% \includegraphics[width=.45\textwidth]{figs/CeBe.eps}
% \vspace{-5mm}
% \caption{The relationship between EB and EC as a function of the QoS-exponent $\theta$\cite{Tang07}, which shows that the solution of the achievable rate and QoS-exponent pair ($\delta, \theta^*$) satisfies $\mathcal{C}_{\rm e}(\theta^*)=\mathcal{B}_{\rm e}(\theta^*) = \delta$. \colr{\bf LHS, a  partir da eq.(9) e (12), gerar um gráfico no MatLab, utilizando a nomenclatura adotada no nosso paper. Use A(t) e R(t) hipotéticos.}}
% \label{fig:CeBe}
% \end{figure}

It is worth noting that the EC characterizes the attainable performance in the large-queue-length regime. By contrast, if the maximum tolerable queue length is finite and short, the maximum supported arrival rates $\check{\delta}$ will be smaller than that predicted by the EC. In such cases, packet loss events occur when the queue is full. As a result, packet retransmission may be required. Hence, systems having a limited queue length in general require more energy. On the other hand, the large-queue-length regime may be regarded as a fundamental limit that can be used as an important benchmark of buffer-aided wireless transmission systems \cite{Qiao09}. 

Finally, in general the derivation of an analytical expression for the EC of an arbitrary stochastic service process remains an open challenge. However, when the service process can be characterized by an independent identically distributed (i.i.d.) process, the EC expression will be substantially simplified \cite{Ahn10}.
%----------------------------------
\subsection{EC of OFDMA Systems}
%----------------------------------
Using the result concerning $\frac{\Lambda_C(-\theta)}{\theta}$ in \cite{Chang94} and \cite[Sec. 7.2]{Chang95}, the EC of a given statistical delay-QoS constraint $\theta$ was analyzed for a simple \textsc{on-off} communication channel in \cite{Qiao09}. Herein, the analysis is extended to realistic \textsc{ofdma} communication channels.
% 
% \noindent \underline{\textsc{on-off} channel} was analyzed in \cite{Qiao09}, considering the EC normalized by the frame duration $T_f$ and bandwidth $B$, or equivalently spectral efficiency in $\left[\frac{\rm bits}{\rm s \cdot Hz}\right]$ for a given statistical delay-QoS  constraint $\theta$:
% %
% \begin{eqnarray}
% \widetilde{\mathcal{C}}_{\rm e}(\textsc{snr}, \theta) &=& \frac{1}{T_fB} \max_{\mathtt{r}\geq 0}\,\, -\frac{\Lambda(-\theta)}{\theta} \qquad \left[\frac{\rm bits}{\rm s \cdot Hz}\right] \nonumber\\
% 	&=& \max_{\mathtt{r}\geq 0}\,\, - \frac{1}{\theta T_fB} \ln(p_{11} + p_{22} e^{-\theta T_f \mathtt{r}}) \nonumber
% \end{eqnarray}
% where \textsc{snr} is the signal-to-noise ratio of the \textsc{on-off}f channel, $\mathtt{r}$ is the transmission rate in [bits/s], and $p_{11}$ and $p_{22}$ are, respectively, the probability of staying in the \textsc{off} and \textsc{on} state, and $p_{11} + p_{22} = 1$.\\
% 
%\noindent \underline{\textsc{ofdma} model}. 

Let the sequence $\{R[i], 1,2,\ldots\}$ be a statistically uncorrelated process. Then, $\mathcal{R}_k$ can be invoked for representing the total amount of data bits delivered on the subcarriers occupied by user $k$ within each frame-duration $T_f$ [second], i.e. we have $\mathcal{R}_k = \sum_{n=1}^N \phi_{k,n} r_{k,n}$, where $\phi_{k,n}\in\{1,\, 0\}$ indicates whether the $n$th subcarrier is assigned to user $k$ or not, and $r_{k,n}$, as defined formally in \eqref{eq:rate_k_n}, is the number of bits per frame-duration $T_f$. Furthermore, a feasible subcarrier assignment indicator matrix ($K\times N$ dimension) should satisfy:
\begin{equation}\label{eq:subcarrier_assign_index}
\pmb \phi \in \Phi \eqdef \left\{ [\phi_{k,n}]_{K\times N} \in \{0,1\} \, \left| \, \sum_{k=1}^K \phi_{k,n}\leq 1 \right. \right\},
\end{equation}
where $K$ is the number of OFDMA users and $N$ is the number of orthogonal subcarriers. The condition (\ref{eq:subcarrier_assign_index}) indicates that at most only a single user is allowed to activate the $n$th subcarrier.

Hence, for the $k$th user, the EC corresponding to an \textsc{ofdma} frame-duration can be formulated as:
\begin{eqnarray}\label{eq:efcap_uncor_ofdma}
&&\mathcal{C}^k_{\rm e}({\bf p}_k, {\pmb \phi}_k, \theta_k) = -\frac{1}{\theta_k} \ln \mathbb{E}\left(e^{-\theta_k \mathcal{R}_k}\right)\nonumber\\
&&= -\frac{1}{\theta_k} \ln \mathbb{E}\left(e^{-\theta_k \sum_{n=1}^N \phi_{k,n} r_{k,n}}\right) \nonumber\\
&&= -\frac{1}{\theta_k} \sum_{n=1}^N  \ln \mathbb{E}\left(e^{-\theta_k \phi_{k,n} r_{k,n}}\right) \nonumber \\
&&= -\frac{1}{\theta_k} \sum_{n=1}^N \phi_{k,n} \ln \mathbb{E}\left(e^{-\theta_k r_{k,n}}\right)  \\
&&= -\frac{1}{\theta_k} \sum_{n=1}^N \phi_{k,n} \ln \mathbb{E}\left(e^{-\theta_k T_f B \log_2 \left( 1 +  \frac{p_{k,n}g_{k,n}}{N_0B}\right)}\right) \nonumber,
\end{eqnarray}
where ${\bf p}_k = [ p_{k,1},\cdots, p_{k,n}, \cdots, p_{k,N}] $ is the $k$th row of the power allocation matrix $\bf P$ defined in \eqref{eq:TxPower}, while ${\pmb \phi}_k = [ \phi_{k,1},\cdots, \phi_{k,n}, \cdots, \phi_{k,N}]$ is the $k$th row of the subcarrier assignment indicator matrix $\pmb \phi$. Furthermore, $p_{k,n}$ and $g_{k,n}$ respectively represent the transmit power and the channel-power-gain on the $n$th subcarrier, which is used for transmission to the $k$th user, with $N_0$ being the single-sided noise-power spectral density and $B$ the bandwidth of a single OFDM subcarrier. The maximum instantaneous transmission rate for the $k$th user on the $n$th subcarrier in a single frame with duration $T_f$ is:
\begin{equation}\label{eq:rate_k_n}
	r_{k,n} = T_fB \log_2\left(1+\frac{g_{k,n}p_{k,n}}{N_0 B}\right)   \qquad \left[\frac{\rm bits}{ T_f}\right].
\end{equation}
Hereafter we assume that the statistical distribution of the channel-power-gain $g_{k,n}$ is known at the transmitter side. Therefore, the probability density distribution (pdf) of $g_{k,n}$, namely $f(g_{k,n})$ is also known at the transmitter side. Furthermore, herein $f(g_{k,n})$ is assumed to be continuously differentiable with respect to $g_{k,n}$. Hence, the expected value in (\ref{eq:efcap_uncor_ofdma}) may be computed as:
\begin{eqnarray}\label{eq:expect}
\mathcal{I}_{k,n} &=& \mathbb{E}\left(e^{-\theta_k r_{k,n}}\right)  \\
&=&\int_0^{\infty} f(g_{k,n})    e^{-\theta_k T_fB \log_2\left(1+ \frac{g_{k,n} p_{k,n}}{N_0 B}\right)} d g_{k,n}. \nonumber
\end{eqnarray}

%-----------------------------
\subsection{OFDMA Power Consumption Model}\label{ofdma_power_consumption_model}
%----------------------------
In order to deal with the RA strategy of energy-efficient communication systems, every single term of the OFDMA system's power consumption must be taken into account, when formulating the optimization objective function. Herein, the total power consumption, which includes a static term and two dynamic terms, is expressed as
\begin{eqnarray}\label{eq:3terms_power_consumpton}
P_\textsc{t}({\pmb \phi}, \mathbf{R}, \mathbf{P}) &=& P_\textsc{cs}\,\, + \,\,\, \underbrace{\varrho\,  \sum_{k=1}^{K} \sum_{n=1}^{N} \phi_{k,n} \,  p_{k,n}}_\text{Power amplifier} \,\,\,  \nonumber \\
&&+ \, \underbrace{\beta \, \sum_{k=1}^{K} \sum_{n=1}^{N} \phi_{k,n} \,  r_{k,n} }_\text{Linear sum-rate dependent power},
\end{eqnarray}
where $\mathbf{R}$ represents the data rate, while $P_\textsc{cs}$ is the static circuit power consumption of electronic devices such as mixers, filters and digital-to-analog converters. The second term is associated with the power consumption of the radio frequency (RF) power amplifier (PA), where $\varrho$ is the PA inefficiency. The third term in \eqref{eq:3terms_power_consumpton} represents a linear sum-rate dependent power dissipation, where the value of $\beta \geq 0$ reflects the relative importance of this term. Depending on the specific values of $\beta$, the third term may represent the baseband back-end signal-processing power dissipation of the transmitter only, of the receivers only, or of both the transmitter and receivers \cite{Ng_Schober12}. Note that herein a linear relationship between the data rate and the signal-processing power consumption has been assumed. These three terms associated with the total power consumption are detailed below.

The total transmit power of a base station (BS) must be bounded and be nonnegative for any feasible power allocation policy. The corresponding power allocation matrix is described by:
\begin{equation}\label{eq:TxPower}
\mathbf{P} \in \bm{\wp} \eqdef \left\{ [p_{k,n}]_{K\times N} \in \mathbb{R}_+ \, \left| \, \sum_{k=1}^K \sum_{n=1}^N p_{k,n} \leq P_{\max} \right. \right\}, 
\end{equation}
where $P_{\max}$ represents the maximum total transmit power available at the BS's transmitter, while the instantaneous power $p_{k,n}$ transmitted on the $n$th subcarrier for the $k$th user can be mapped into the maximum instantaneous transmission rate $r_{k,n}$. More specifically, from \eqref{eq:rate_k_n} we obtain:
\begin{equation}\label{eq:p_k_n}
p_{k,n} \geq  \frac{N_0B\left(2^{r_{k,n}/T_fB} - 1\right)}{g_{k,n}}   \qquad \text{[W]}.
\end{equation}
Furthermore, the static power consumption of the circuitry, namely $P_\textsc{cs}$ in \eqref{eq:3terms_power_consumpton}, is determined by the active circuit blocks, such as the analog-to-digital converter (\textsc{adc}), digital-to-analog converter (\textsc{dac}), synthesizer (syn), mixer ({mix}), low power amplifier (\textsc{lpa}), intermediate frequency amplifier (\textsc{ifa}) as well as the transmitter and receiver filters (filt, filr) \cite{Cui05}. Hence, the static power consumption of the circuitry can be decomposed into several terms as follows:
$$
P_\textsc{cs} = 2P_{\rm syn} + P_{\rm mix} + P_{\textsc{lpa}} + P_{\rm filt} + P_{\rm filr} + P_{\textsc{ifa}} + P_{\textsc{adc}}.
$$
%
% It is a common assumption in microelectronics that the chip power dissipation can be satisfactorily modeled as the sum of a \emph{static term} and a \emph{dynamic term}: \textcolor{red}{is it necessary to talk about chip power dissipation here?}
% $$
% P_{\rm chip} = V_{\rm dd} \, I_{\rm leak} + a \, f_{\rm ck} \, C_{\rm ap} \, V_{\rm dd}^2\qquad [W],
% $$
% where constant $a$ is related to the effective fraction of gates switching, $f_{\rm ck}$ is the clock frequency, and $C_{\rm ap}$ is the circuit capacitance. Additionally, $V_{dd}$ is the DC voltage source, and $I_{\rm leak}$ is the leakage current. \textcolor{red}{If the frequency is dynamically scaled with the information rate $r$, it is reasonable to model the power dissipation as a linear function of the rate with a constant offset, as discussed in \cite{Schurgers01}. Therefore, as pointed out above, the power dissipation in the RF front-end during transmission is dominated by the \textsc{mix}, syn, filter and \textsc{dac} blocks consumption, and can be well modeled as a rate-independent constant \cite{Cui05}.} \textcolor{blue}{can't see any connection between this paragraph with the above description} 
%
As a result, the overall power consumption at the BS, namely \eqref{eq:3terms_power_consumpton}, may be reformulated as:
\begin{equation}\label{eq:reformulated_total_power}
	P_\textsc{t}({\pmb \phi},\mathbf{R}, \mathbf{P}) = \varrho P({\pmb \phi}, \mathbf{P}) \,\, +\,\, P_\textsc{c}({\pmb \phi}, \mathbf{R}),
\end{equation}
where the PA inefficiency $\varrho$ is expressed as $\varrho = \left(\frac{\textsc{papr}}{\xi} - 1\right)$  \cite{Cui05}, with the numerator being the peak-to-average power ratio (PAPR) and  $\xi$ the drain efficiency of the PA. The parameter \textsc{papr} depends on the specific modulation scheme.  
Explicitly, the circuit power $P_\textsc{c}({\pmb \phi},\mathbf{R})$ is modeled as a function of the data rate and the subcarrier allocation policy, yielding
\begin{eqnarray}\label{eq:2terms_pc}
P_\textsc{c}({\pmb \phi},\mathbf{R}) &=& P_\textsc{cs}\, + \underbrace{\beta \, \sum_{k=1}^{K} \sum_{n=1}^{N} \phi_{k,n} \,  r_{k,n} }_\text{Linear sum-rate dependent power},  
\end{eqnarray}
which contains a static term and a dynamic term, corroborating the power consumption model of \eqref{eq:3terms_power_consumpton}. 

Observe that the last term in \eqref{eq:2terms_pc} represents a second-order effect, which leads to slowly increasing values as the information rate increases. As a result, $P_\textsc{cs}$ of \eqref{eq:2terms_pc} becomes dominant. Hence,  for the sake of simplicity, in this paper a constant circuitry power consumption model has been assumed, i.e. $P_\textsc{c}({\pmb \phi},\mathbf{R}) \approx P_\textsc{cs} =$ constant, implying $\beta=0$.

It is worth noting that in this paper we mainly aim for maximizing the EEE subject to a given delay-QoS constraint of a realistic OFDMA network. Note that the EC $\mathcal{C}_{\rm e}(\theta)$ can be considered as the maximal throughput per frame-duration under the QoS-exponent $\theta$. Therefore, by interpreting $\theta$ as the delay-QoS constraint, it is possible to formulate an equivalent problem, which aims for maximizing the EC for a given statistical delay-QoS constraint. As a result, we can further maximize the EEE, which can be simply formulated as the ratio of the EC to the total network's energy consumption, in $\left[\frac{\rm bits}{\rm Joule}\right]$. In Sec. \ref{sec:probl_formul} and Sec. \ref{sec:solution}, we will focus our attention on the problem formulation, as well as on designing the corresponding iterative RA algorithms, respectively.

%=============================
\section{Formulation of the Downlink OFDMA EEE Maximization Problem} \label{sec:probl_formul}
%=============================
In this paper, the downlink of an OFDMA network having $N$ subcarriers and a total bandwidth of $NB$ is considered. As shown in \eqref{eq:large_queue_length} and \eqref{eq:queue_length_small}, since the approach adopted is based on asymptotic analysis, the buffers at the BS are assumed to be large enough and always full, so that no empty scheduling slot is caused by having insufficient source packets in the buffers.
\subsection{The Original EEE-Maximization Problem}
Before presenting our EEE-optimal design, let us formally define the EEE for the downlink OFDMA network as the ratio of the overall EC to the total consumed energy in [bits/Joule]:
\begin{eqnarray}\label{eq:ee_def}
\eta_{\textsc{e}}({\bm \theta, \pmb  \phi, \bf P}) &\eqdef& \frac{\mathcal{C}_{\rm e}(\bm \theta,\pmb\phi,{\bf P})}{T_f P_\textsc{t}(\pmb \phi, {\bf R}, {\bf P})}  \, 
= \, \frac{\mathcal{C}_{\rm e}(\bm \theta,\pmb\phi,{\bf P})}{\mathcal{U}_\textsc{p}(\pmb\phi, {\bf R},{\bf P})}
\end{eqnarray}
\begin{equation*}
=-\frac{1}{T_f (\varrho P + P_c) }\sum_{k=1}^K  \sum_{n=1}^N  \frac{\phi_{k,n}}{\theta_k} \, \ln \mathbb{E}\left(e^{-\theta_k  r_{k,n}}\right) ,
\end{equation*}
where $P_\textsc{t}(\pmb \phi, {\bf R}, {\bf P})$ is given by \eqref{eq:3terms_power_consumpton} and \eqref{eq:reformulated_total_power}. Note that the EEE definition of \eqref{eq:ee_def} considers the delay-QoS requirements specified by $\bm \theta$. In this definition, the EEE is described as a delay-QoS-guaranteed metric. Hence, our EEE-optimal design conceived for the downlink of OFDMA systems can be formulated as the EEE maximization under statistical delay-QoS guarantees according to:
{
\begin{eqnarray}\label{eq:eee_opt}
\eta_{\textsc{e}}^{\rm opt}{( \bm \theta)}	&=& \underset{\pmb{\phi} \in \bm{\Phi}, \,\, {\bf P} \in {\bm\wp}}{\rm maximize} \quad \eta_{\textsc{e}}(\bm \theta, \pmb  \phi, \bf P)\\
\text{s.t.}	\quad \mbox{C1:} & & \mathcal{C}^k_{\rm e}({\bf p}_k, {\pmb \phi}_k, \theta_k)\geq \mathcal{C}^{k,\min}_{\rm e}, \quad \forall \, k \nonumber\\
\textsc{C2:} & & \sum_{k=1}^{K} \sum_{n=1}^{N} \phi_{k,n} \, p_{k,n} \leq P_{\max}, \quad p_{k,n} \geq 0 \nonumber \\
\textsc{C3:} & & \sum_{k=1}^K \phi_{k,n}\leq 1, \nonumber\\
\textsc{C4:} & & \phi_{k,n} \in \{0,1\}, \nonumber\\
\textsc{C5:} & & \sum_{k=1}^{K}N_k = N, \quad N_k \in \mathbb{Z}_+ \nonumber. 
\end{eqnarray}}

Constraint $\mbox{C1}$ holds for the minimum EC that the $k$th user should achieve. $\mbox{C2}$ ensures that the total power allocated to the $N$ subcarriers of $K$ users does not exceed the maximum transmit power  $P_{\max}$ available at the BS. Constraints $\mbox{C3}$ and $\mbox{C4}$  are imposed in order to guarantee that each subcarrier is used at most by one user, hence avoiding inter-user interference. The feasible region for the optimization variables $\pmb{\phi}$ and ${\bf P}$ is described by the constraints $\mbox{C1}-\mbox{C5}$.

Additionally, at a given time instant, the channel-power-gains of the different OFDMA subcarriers belonging to a specific user, for example $g_{k,n}, \,\, n=1\ldots N$ for the $k$th user,  may be modelled by independent identically distributed (i.i.d.) random variables.   As a result,  we can simply use $f(g_{k,n})$ or $f_{k,n}, \,\, n=1\ldots N$ to represent the pdf of the channel-power-gain on each subcarrier. 

The EEE optimization problem \eqref{eq:eee_opt} can be classified as a nonlinear fractional program\cite{Dinkelbach67,Schaible76b}, whose  objective function is the ratio of two functions and it is generally a non-convex (non-concave) function. In the following, we will show that the EC of Rayleigh fading channels (RFC) is a concave function, while the EEE function is quasi-concave, which is consistent with the above statement. More specifically, the numerator of the objective function of \eqref{eq:eee_opt} is concave with respect to (w.r.t.) the variables $\phi_{k,n}$ and $p_{k,n}$, since it is the non-negative sum of multiple concave functions. Furthermore, the denominator is affine, i.e. convex as well as concave. It is well known that for this kind of objective function, the problem is quasi-concave \cite{Schaible76}. The proof of these properties is offered in Lemma \ref{lemma:2} and Appendix \ref{app:1}. 

%==================================================================
\subsection{Relaxations of the EEE-Optimal Design}\label{sec:relax_EEE_ESE}
%==================================================================
In order to conceive an EEE-optimal design we have to solve Problem \eqref{eq:eee_opt} to find the optimal subcarrier and power allocation. In fact, the subcarrier allocation itself is a combinatorial integer programming problem, which is in general NP-hard. Hence, introducing a relaxation into the subcarrier constraints makes Problem \eqref{eq:eee_opt} more tractable. 
The approach adopted herein for the mixed-integer programming problem\footnote{Problem \eqref{eq:eee_opt} is a mixed-integer programming problem, because the subcarrier allocation variables are discrete, while the power allocation variables are continuous.} of \eqref{eq:eee_opt} relies on approximating the integer part of Problem \eqref{eq:eee_opt} by its continuous relaxation, since in general continuous-variable based optimization problems are easier to solve than discrete-variable based combinatorial optimization problems. The idea of continuous relaxation is to enlarge the feasible set, while making sure that it includes, but is not limited to, all feasible solutions that satisfy the original constraints\cite{Cioffi02, Wong99}. Therefore, instead of forcing the optimization variable (subcarrier occupancy indicator) to be either 0 or $1$, the constraint (C4) in Problem \eqref{eq:eee_opt} can be relaxed to $0 \leq \phi_{k,n} \leq 1$, or equivalently to $0 \leq \phi_{k,n} B \leq B$ and  $0 \leq \phi_{k,n} p_{k,n} \leq p_{k,n}$.

The relaxation of the subcarrier assignment variables, allowing them to take continuous values over the $[0,\, 1]$ interval, is equivalent to the multi-user time-sharing of each subcarrier over a large number of OFDM symbols\cite{Wong99, Yu06, Kent_2013:EEM_fractional_programming} and generally does not solve exactly the original problem. Wireless communication channels are typically time varying and the channels may not stay unchanged long enough for time-sharing to be feasible \cite{Wong99}. Fortunately, it has been shown that the solution of the relaxed problem under the time-sharing condition is arbitrarily close to the solution of the original problem, when the number of subcarriers tends to infinity \cite{Yu06}. In fact, the gap between the two solutions can be small even for a small number of subcarriers \cite{cioffi06,Yu06, Kent_2013:EEM_fractional_programming}.

Hence, this relaxation is applied to the subcarrier assignment indicator set of \eqref{eq:subcarrier_assign_index}, to the power allocation set of \eqref{eq:TxPower}, to the achievable rate of \eqref{eq:rate_k_n} and to the overall OFDMA EC of \eqref{eq:efcap_uncor_ofdma}, respectively as follows:
\begin{eqnarray}\label{eq:subcarrier_assign_index_relax}
\underline{\pmb \phi} \in \underline{\bm\Phi} & \eqdef & \left\{  \underline{\pmb\phi} \in [0,1]^{K\times N} \, \left| \,  \sum_{k=1}^K \underline{\phi}_{k,n} \leq 1 \right. \right\}, \\
\label{eq:TxPower_relax}
\underline{\mathbf{P}} \in \underline{\bm{\wp}} & \eqdef & \left\{ \underline{\bf P}  \in \mathbb{R}^{K\times N}_+  \, \left|\,\sum_{k=1}^K \sum_{n=1}^N \underline{p}_{k,n} \leq P_{\max} \right. \right\},\\
\label{eq:rate_k_n_relax}
\underline{r}_{k,n} & = & T_fB \log_2\left(1+\frac{g_{k,n}\underline{p}_{k,n}}{\underline{B}_{k,n} N_0}\right)  ,\\
\label{eq:efcap_overall_relax}
\underline{\mathcal{C}_{\rm e}}({\bm \theta}, \underline{\pmb \phi}, \underline{\mathbf{P}}) & = &  - \sum_{k=1}^K \sum_{n=1}^N \frac{\phi_{k,n}}{\theta_k} \ln \mathbb{E}\left(e^{-\theta_k \underline{r}_{k,n}}\right),
\end{eqnarray}
where the new subcarrier assignment index $\underline{\phi}_{k,n}$ is a continuous variable in the interval $[0,1]$, and it can be interpreted as the portion of subcarrier $n$ assigned to user $k$, i.e. we have $\underline{B}_{k,n} = \underline{\phi}_{k,n} B$ \cite{Cioffi00, Cioffi00_conf, Cioffi02}, or interpreted as the time-sharing factor of subcarrier assignment\cite{Kent_2013:EEM_fractional_programming}. Hence, instead of restricting the boundaries of the partitions between the two users to align with the bin boundaries as the integer programming does, in \cite{Cioffi00_conf} the boundary is allowed to be anywhere in the bin, hence relaxing the integer programming problem into a continuous-variable optimization problem.

As a beneficial result of the  $\pmb \phi$-relaxation, the following variable transformations can be introduced: $\underline{B}_{k,n} = \underline{\phi}_{k,n} B$ and  $\underline{p}_{k,n} = \underline{\phi}_{k,n} p_{k,n}$ for all $\underline{\phi}_{k,n}\in[0,\,1]$. Then, a modified version of the original EEE-maximization problem of \eqref{eq:eee_opt} may be obtained as: 

\begin{eqnarray}\label{eq:eee_opt_relax}
\underline{\eta_{\textsc{e}}}^{\rm opt}{( \bm \theta)}	&=& \underset{\underline{\pmb{\phi}} \in \underline{\bm{\Phi}}, \,\, \underline{\mathbf{P}} \in \underline{\bm\wp}}{\rm maximize} \quad \eta_{\textsc{e}}(\bm \theta, \underline{\pmb\phi}, \underline{\mathbf{P}}) \\
\text{s.t.}	\quad \mbox{C1:} & & \underline{\mathcal{C}^k_{\rm e}}({\bf p}_k, {\pmb \phi}_k, \theta_k) \geq \mathcal{C}^{k,\min}_{\rm e}, \quad \forall \, k\nonumber\\
\mbox{C2:} & & \sum_{k=1}^{K} \sum_{n=1}^{N} \underline{p}_{k,n} \leq P_{\max}, \quad \underline{p}_{k,n} \geq 0 \nonumber \\
\mbox{C3:} & & \sum_{k=1}^K \underline{\phi}_{k,n}\leq 1, \nonumber\\
\mbox{C4:} & & \underline{\phi}_{k,n} \in [0,1], \nonumber\\
\mbox{C5:} & & \sum_{k=1}^{K}N_k = N, \quad N_k \in \mathbb{R}_+ .\nonumber 
\end{eqnarray}

%%%%%%%%%%%%%%%%%%%%%%%%%%%%%%%%%%%%%%%%%%%%%%%%%%%%%%%%%%%%%%%%%%%%%%%%%%%%%%%%%%%
\subsection{Calculation of the EC for NLOS Rayleigh Fading Channels}
%%%%%%%%%%%%%%%%%%%%%%%%%%%%%%%%%%%%%%%%%%%%%%%%%%%%%%%%%%%%%%%%%%%%%%%%%%%%%%%%%%%

When a non-line-of-sight (NLOS) Rayleigh fading propagation channel is considered, the channel-power-gain $g_{k,n}$ is an exponentially distributed random variable. As a result, the expectation in  \eqref{eq:efcap_uncor_ofdma} is readily obtained by:
\begin{eqnarray}\label{eq:expect_exp}
\mathcal{I}_{k,n} &=& \mathbb{E}\left(e^{-\theta_k r_{k,n}}\right)   \\
&=& \int_0^{\infty} e^{-\theta_k T_fB \log_2\left(1+ \frac{g_{k,n} p_{k,n}}{N_0 B}\right)} f(g_{k,n}) \,\, d g_{k,n} \nonumber \\
&=& \int_0^{\infty} \left(1+ \frac{g_{k,n} p_{k,n}}{N_0 B}\right)^{\frac{-\theta_k T_fB}{\ln(2)}} \,\, \ell e^{-\ell g_{k,n}} \, d g_{k,n}. \nonumber
\end{eqnarray}
Employing the following substitutions:
\[
 t = 1 \, + \, g_{k,n} D, \qquad D = \dfrac{p_{k,n}}{N_0 B}, \qquad A_k = \dfrac{\theta_k T_f B}{\ln(2)}
\]
and assuming $\ell = 1$, while $A_k, D>0$, the integral may be calculated as:
\begin{eqnarray}\label{eq:expect_exp2}
 && \int_0^\infty e^{-g_{k,n}} (1+g_{k,n}D)^{-A_k} \,\, dg_{k,n} \nonumber \\
 &=& \dfrac{e^{\frac{1}{D}}}{D} \int_1^\infty e^{-\frac{t}{D}} t^{-A_k} dt \nonumber \\
 &=& \dfrac{e^{\frac{1}{D}}}{D} \, \, E_{A_k}\left(\frac{1}{D}\right),
\end{eqnarray}
where $E_n(x)$ is the exponential integral function. From \eqref{eq:efcap_uncor_ofdma}, \eqref{eq:expect_exp} and \eqref{eq:expect_exp2} the EC of the $k$th user can be calculated for a RFC as:
\begin{eqnarray}\label{eq:rfcEC}
 &&\mathcal{C}_{e,\texttt{RFC}}^k(\theta_k, {\pmb \phi}_k, {\bf p}_k) = \\
 &&=-\frac{1}{\theta_k} \sum_{n=1}^N \phi_{k,n} \, \ln\left( \frac{N_0 B}{p_{k,n}} e^{\frac{N_0 B}{p_{k,n}}} E_{A_k} \left( \frac{N_0 B}{p_{k,n}}\right) \right)\nonumber,
\end{eqnarray}
and the system's total EC is written as:
\begin{eqnarray}\label{eq:rfcTEC}
 &&\mathcal{C}_{e,\texttt{RFC}}({\bm \theta}, {\pmb \phi}, {\bf P})  = \\
 &&\sum_{k=1}^K \sum_{n=1}^N -\frac{ \phi_{k,n}}{\theta_k}  \, \ln\left( \frac{N_0 B}{p_{k,n}} e^{\frac{N_0 B}{p_{k,n}}} E_{A_k} \left( \frac{N_0 B}{p_{k,n}}\right) \right), \nonumber
\end{eqnarray}
while the relaxed form of \eqref{eq:rfcTEC} may be directly defined as:
\begin{equation}\label{eq:rfcTECR}
 \underline{\mathcal{C}}_{e,\texttt{RFC}}({\bm \theta}, \underline{\pmb \phi}, \underline{\bf P})  = \sum_{k=1}^K \sum_{n=1}^N -\frac{ \underline{\phi}_{k,n}}{\theta_k}  \, \ln\left(\underline{\mathcal{I}}^{\texttt{RFC}}_{k,n} \right),
\end{equation}
where
\begin{equation}\label{eq:rfcTECR_I}
 \underline{\mathcal{I}}^{\texttt{RFC}}_{k,n} = \frac{N_0 \underline{B}_{k,n}}{\underline{p}_{k,n}} e^{\frac{N_0 \underline{B}_{k,n}}{\underline{p}_{k,n}}} E_{A_k} \left[ \frac{N_0 \underline{B}_{k,n}}{\underline{p}_{k,n}}\right].
\end{equation}

The concavity of the system's EC is discussed in the Proof of Lemma \ref{lemma:1}.
\begin{lemma}\label{lemma:1}
\it For NLOS Rayleigh fading channels, the relaxed EC function \eqref{eq:rfcTECR} of the system is concave in both $\underline{p}_{k,n}$ and $\underline{\phi}_{k,n}$.
\end{lemma}
%%%%%%%%%%%%%%%%%%%%%%%%%%%%%%%
\begin{IEEEproof}
 See Appendix \ref{app:1}.
\end{IEEEproof}
%%%%%%%%%%%%%%%%%%%%%%%%%%%%%%%

%%%%%%%%%%%%%%%%%%%%%%%%%%%%%%%%%%%%%%%%%%%%%%%%%%%%%%%%%%%%%%%%%%%%%%%
\section{An Algorithm for Solving the OFDMA EEE-Maximization Problem} \label{sec:solution}
%%%%%%%%%%%%%%%%%%%%%%%%%%%%%%%%%%%%%%%%%%%%%%%%%%%%%%%%%%%%%%%%%
The energy efficiency of wireless networks may be defined as the number of transmitted bits per unit of energy [Joule]. Hence, given the EC defined for Rayleigh fading channels in (\ref{eq:rfcTEC}), we may define the system's EEE in [bits/Joule] as:
\begin{eqnarray}\label{eq:eeeRFC}
 &&\eta_E^{\texttt{RFC}} \triangleq \dfrac{\mathrm{C}^{\texttt{RFC}}_e({\bm \theta},{\pmb \phi},{\bf P})}{T_f P_\textsc{t}({\pmb \phi},\mathbf{R},\mathbf{P})} \\
 &&\triangleq \dfrac{\sum\limits_{k=1}^K \sum\limits_{n=1}^N -\dfrac{\phi_{k,n}}{\theta_k} \, \ln\left(\frac{N_0 B}{p_{k,n}}\, e^{\frac{N_0 B}{p_{k,n}}} \, E_{A_k}\left(\frac{N_0 B}{p_{k,n}}\right)\right)}{T_f \, \left(P_\textsc{c} \, + \, \varrho \, \sum\limits_{k=1}^K \sum\limits_{n=1}^N \phi_{k,n} p_{k,n}\right)} \nonumber,
\end{eqnarray}
\noindent where again, $\varrho$ is the PA inefficiency and $P_\textsc{c}$ the circuitry power dissipation at the BS.
Therefore, the EEE optimization problem of OFDMA systems operating in NLOS Rayleigh fading channels under a specific statistical delay-QoS provision is formulated as:
\begin{eqnarray}\label{prob:EEE_M}
\mbox{maximize} &  & \eta_E^{\texttt{RFC}} \\
\text{s.t.}	\quad \mbox{C1: } &  & \mathcal{C}_{e,\texttt{RFC}}^k(\theta_k, {\pmb \phi}_k, {\bf p}_k)\geq\mathcal{C}_{e}^{k,\min}, \nonumber \\ 
\mbox{C2: } &  & \sum_{k=1}^{K}\sum_{n=1}^{N} \phi_{k,n} \, {p}_{k,n}\leq P_{\max}, \nonumber \\
\mbox{C3: } &  & \sum_{k=1}^{K}\phi_{k,n}\leq1\mbox{, }\forall n \nonumber \\
\mbox{C4: } &  & {p}_{k,n}\in\mathbb{R}_{+}\mbox{, }\forall k,n \nonumber \\
\mbox{C5: } &  & \phi_{k,n}\in[0,1]\mbox{, }\forall k,n \nonumber.
\end{eqnarray}
Observe in \eqref{eq:eeeRFC} that the EEE is the ratio of a nonnegative weighted sum of concave functions over a nonnegative affine function. Therefore, the following Lemma holds:

\begin{lemma}\label{lemma:2}
 The EEE function $\eta_E^{\texttt{RFC}}$ of \eqref{eq:eeeRFC} is quasi-concave.
 \begin{IEEEproof}
  From Lemma \ref{lemma:1} we infer that $\eta_E^{\texttt{RFC}}$ is the ratio of a concave function to an affine positive function. According to \cite[Table 5.5 on P. 165]{avriel} this ratio results in a semi-strictly quasi-concave function.
 \end{IEEEproof}
\end{lemma}

Therefore, the EEE optimization problem  \eqref{prob:EEE_M}  and its relaxed form relying on \eqref{eq:rfcTECR}-\eqref{eq:rfcTECR_I} are concave fractional programming problems, whose objective functions are cast in a fractional form. 
%
%
% \begin{eqnarray*}
% \mbox{maximize} &  & \xi_e =\dfrac{f({\bf P},{\bm \phi})}{u({\bf P},{\bm \phi})} \\
% \mbox{subject to}\\
% \mbox{C1: } &  & \mathcal{C}_{e,\texttt{RFC}}^k(\theta_k, {\bm \phi}_k, {\bf p}_k)\geq\mathcal{C}_{e}^{k,\min}\\
% \mbox{C2: } &  & \sum_{k=1}^{K}\sum_{n=1}^{N} \phi_{k,n} \, {p}_{k,n}\leq P_{\max},\\
% \mbox{C3: } &  & \sum_{k=1}^{K}\phi_{k,n}\leq1\mbox{, }\forall n,\\
% \mbox{C4: } &  & {p}_{k,n}\in\mathbb{R}_{+}\mbox{, }\forall k,n,\\
% \mbox{C5: } &  & \phi_{k,n}\in[0,1]\mbox{, }\forall k,n.
% \end{eqnarray*}
In order to solve the above fractional programming problem, Dinkelbach's classic method\cite{Dinkelbach67,Schaible76b} may be invoked.

%-------------------------------
\subsection{Dinkelbach's Method}
%-------------------------------
Since concave-convex fractional programs share important properties with concave optimization problems, it is possible to solve concave-convex fractional programs with the aid of standard methods developed for concave optimization problems. Here, we use Dinkelbach's method \cite{Dinkelbach67,Schaible76b}, which operates in an inner-outer iteration manner.

Upon using Dinkelbach's iterative method \cite{Dinkelbach67,Schaible76b}, the quasi-concave problem posed in \eqref{eq:eee_opt} can be solved in a parameterized concave form. To elaborate a little further, the original concave-convex fractional program has a form similar to 
$$
\underset{x\in \mathcal{F}}{\rm maximize} \quad q(x) = \frac{f(x)}{z(x)},
$$
where $\mathcal{F}$ is a compact, connected set and $z(x)>0$ is assumed. For the sake of notational simplicity, we define $\mathcal{F} \supset \{ \bm{\Phi}, \, \bm{\wp} \} $ as the set of feasible solutions of the original optimization problem described by \eqref{eq:eee_opt}. The original problem can be associated with the following parametric concave problem \cite{Dinkelbach67,Schaible76}:
$$
\underset{x\in \mathcal{F}}{\rm maximize} \quad  f(x) - q\, z(x),
$$
where $q\in \mathbb{R}$ is treated as a parameter. The objective function, which is denoted hereafter by $F(q)$ for this parametric problem, is convex, continuous-valued and strictly decreasing. Additionally, without loss of generality, we define the {\it maximum EEE} $q^*$ of the  system considered as:
\begin{equation}\label{eq:ee_max}
q^* =	\frac{\mathcal{C}_{\rm e}(\bm \theta,\pmb{\phi}^*,{\bf P}^*)}{\mathcal{U}_\textsc{p}(\pmb\phi^*,{\bf P}^*)} \, = \, \quad \underset{\pmb{\phi} \in \bm{\Phi}, \,\, {\bf P} \in {\bm\wp}}{\rm maximize} \quad  \frac{\mathcal{C}_{\rm e}(\bm \theta,\pmb{\phi},{\bf P})}{\mathcal{U}_\textsc{p}(\pmb\phi,{\bf P})}.
\end{equation}
It is plausible that we have  
$$
 \qquad \left\{\begin{matrix}
F(q)>0 & \Leftrightarrow & q<q^*\\ 
F(q)=0 & \Leftrightarrow & q=q^*\\ 
F(q)<0 & \Leftrightarrow & q>q^*. 
\end{matrix}\right.
$$
Hence, Dinkelbach's method presented in Algorithm \ref{alg:Dink} solves the following problem:
\begin{equation}\label{eq:dink_solv}
\underset{\pmb{\phi} \in \bm{\Phi}, \,\, {\bf P} \in {\bm\wp}}{\rm maximize} \quad  \mathcal{C}_{\rm e}(\pmb{\phi},{\bf P})  - q \, \mathcal{U}_\textsc{p}(\pmb\phi,{\bf P}),
\end{equation}
which is equivalent to finding the root of the nonlinear equation $F(q)=0$.
\begin{algorithm}[!h]
\small
\caption{Dinkelbach's Method}
Input: \hspace{12mm}  $q_0$ satisfying $F(q_0)\geq0$; tolerance $\epsilon$ \\
Initialization: \hspace{2mm}  $n \leftarrow 0$\\
{\bf repeat}
 \begin{itemize}
  \item[] Solve Problem \eqref{eq:dink_solv} with $q=q_n$ to obtain $\pmb{\phi}^*$ and ${\bf P}^*$;
  \item[] $q_{n+1} \leftarrow  \frac{\mathcal{C}_{\rm e}(\pmb{\phi}^*,{\bf P}^*)} {\mathcal{U}_\textsc{p}(\pmb\phi^*,{\bf P}^*)}$;
  \item[] $n \leftarrow n+1$;
 \end{itemize}
{\bf until} $|F(q_n)|\leq\epsilon$
\label{alg:Dink}
\end{algorithm}

Dinkelbach's method in fact constitutes the application of Newton's method to a nonlinear fractional program \cite{Schaible83}. As a result, the sequence converges to the optimal point at a superlinear convergence rate \cite{Schaible76b}. In summary, Dinkelbach's method \cite{Dinkelbach67} is an iterative technique of finding the increasing values of feasible $q$ by solving the parameterized problem of $\underset{\pmb{\phi},{\bf P}}{\max}\,\, F(q_n) = 
 \underset{\pmb{\phi},{\bf P}}{\max}\,\, \,\, \{\mathcal{C}_{\rm e}(\pmb{\phi},{\bf P})  - q_n \, \mathcal{U}_\textsc{p}(\pmb\phi,{\bf P}) \}$
at the $n$th iteration. This iterative process continues until the absolute difference value  $|F(q_n)|$ becomes less than or equal to a pre-specified tolerance threshold $\epsilon$.

The parametric version of the relaxed EEE-maximization problem of \eqref{eq:eee_opt_relax} is described as:
\begin{eqnarray}\label{eq:DinkEEEmax}
\mbox{maximize} &  & \mathcal{C}_{\rm e}(\pmb{\phi},{\bf P})  - q \, \mathcal{U}_\textsc{p}(\pmb\phi,{\bf P}),\\
\mbox{s.t.} \quad \mbox{C1: } &  & \mathcal{C}_{e,\texttt{RFC}}^k(\theta_k, {\pmb \phi}_k, {\bf p}_k)\geq\mathcal{C}_{e}^{k,\min},\nonumber\\
\mbox{C2: } &  & \sum_{k=1}^{K}\sum_{n=1}^{N}{p}_{k,n}\leq P_{\max},\nonumber\\
\mbox{C3: } &  & \sum_{k=1}^{K}\phi_{k,n}\leq1\mbox{, }\forall n\nonumber\\
\mbox{C4: } &  & {p}_{k,n}\in\mathbb{R}_{+}\mbox{, }\forall k,n\nonumber\\
\mbox{C5: } &  & \phi_{k,n}\in[0,1]\mbox{, }\forall k,n.\nonumber
\end{eqnarray}
Since this is a concave problem and the conditions (C1), (C2) and (C3) satisfy Slater's conditions \cite{slater}, one can solve the dual problem to obtain the primal solution with zero duality gap. Therefore, the Lagrangian over $\mathbf{P}$ and $\pmb \phi$ for the optimization problem of \eqref{eq:DinkEEEmax} is presented in \eqref{eq:lagrangian}.
%%%%%%%%%%%%%%%%%%%%%%%%%%%%%%%%%%% Equation at the bottom of the next page (34)
\begin{figure*}[!b]
\normalsize
\centering
\vspace*{-5mm}
\hrulefill
 \begin{eqnarray}\label{eq:lagrangian}
 \mathcal{L}({\bf P},{\pmb \phi},{\bm \nu},\lambda) &=& \sum\limits_{k=1}^K \sum\limits_{n=1}^N -\dfrac{\phi_{k,n}}{\theta_k}
  \, \ln\left(\frac{N_0 B}{p_{k,n}}\, e^{\frac{N_0 B}{p_{k,n}}} \, E_{A_k}\left(\frac{N_0 B}{p_{k,n}}\right)\right) - \, q_{i-1} \, \left[T_f\left(P_\textsc{c} \, + \, \varrho \, \sum\limits_{k=1}^K \sum\limits_{n=1}^N \phi_{k,n} p_{k,n}\right)\right] \nonumber \\
 && + \sum\limits_{k=1}^K \nu_k \left(\sum\limits_{n=1}^N -\dfrac{\phi_{k,n}}{\theta_k} \ln\left(\frac{N_0 B}{p_{k,n}}\, e^{\frac{N_0 B}{p_{k,n}}} \, E_{A_k}\left(\frac{N_0 B}{p_{k,n}}\right)\right) - \mathcal{C}_e^{k,\min}\right) \nonumber \\
 && + \lambda \left(P_{\max} - \sum\limits_{k=1}^K \sum\limits_{n=1}^N \phi_{k,n} p_{k,n}\right) 
\end{eqnarray}
\end{figure*}

Additionally, the following relationship holds:
\begin{eqnarray}\label{eq:dualrelation}
 && \argmin_{{\bm \nu},\lambda} \quad \sup_{\mathbf{P},{\pmb \phi}} \, \mathcal{L}({\bf P},{\pmb \phi},{\bm \nu},\lambda) \, \equiv \nonumber \\
 && \equiv \, \argmax_{\mathbf{P},{\pmb \phi}} \, \mathcal{C}_{\rm e}(\pmb{\phi},{\bf P})  - q \, \mathcal{U}_\textsc{p}(\pmb\phi,{\bf P}),
\end{eqnarray}
where the Lagrange dual function is given by $\sup \mathcal{L}$, i.e. by the supremum of the Lagrangian. The relationship in \eqref{eq:dualrelation} is further developed in \eqref{eq:developeddual}, which leads to the conclusion that the dual problem of
\begin{equation*}
 \argmin_{{\bm \nu},\lambda} \quad \sup_{\mathbf{P},{\pmb \phi}} \, \mathcal{L}({\bf P},{\pmb \phi},{\bm \nu},\lambda)
\end{equation*}
may be solved by solving $KN$ subproblems of the form presented in \eqref{eq:newform}, while the dual variables $\bm\nu$ and $\lambda$ can be updated by applying the subgradient method of\cite{Boyd04}.
%%%%%%%%%%%%%%%%%%%%%%%%%%%%%%%%%%% Equation at the bottom of the next page (36)
\begin{figure*}[!b]
 \centering
 \normalsize
 \hrulefill
 \vspace*{1mm}
\begin{eqnarray}\label{eq:developeddual}
 &&\sup_{{\bf P},{\pmb \phi}} \, \, \mathcal{L}({\bf P},{\pmb \phi},{\bm \nu},\lambda) = \argmax_{{\bf P},{\pmb \phi}}\sum\limits_{k=1}^K \sum\limits_{n=1}^N -\dfrac{\phi_{k,n}}{\theta_k}
   \, \ln\left(\frac{N_0 B}{p_{k,n}}\, e^{\frac{N_0 B}{p_{k,n}}} \, E_{A_k}\left(\frac{N_0 B}{p_{k,n}}\right)\right) \nonumber \\
 && + \sum\limits_{k=1}^K \sum\limits_{n=1}^N -\dfrac{\nu_k \, \phi_{k,n}}{\theta_k} \ln\left(\frac{N_0 B}{p_{k,n}}\, e^{\frac{N_0 B}{p_{k,n}}} \, E_{A_k}\left(\frac{N_0 B}{p_{k,n}}\right)\right) - \sum\limits_{k=1}^K \nu_k \, \mathcal{C}_e^{k,\min}  \\
  && + \left[\lambda P_{\max} - \lambda \left( \sum\limits_{k=1}^K \sum\limits_{n=1}^N \phi_{k,n} p_{k,n} \right) - \, q_{i-1} \, T_f \, \varrho \, \left( \sum\limits_{k=1}^K \sum\limits_{n=1}^N \phi_{k,n} p_{k,n}\right) - \, q_{i-1} \, T_f \, P_\textsc{c} \right]\nonumber \\
 &&= \argmax_{{\bf P},{\pmb \phi}} \sum\limits_{k=1}^K \sum\limits_{n=1}^N \left[\dfrac{\phi_{k,n}(\nu_k + 1)}{\theta_k}
  \, \ln\left(\frac{N_0 B}{p_{k,n}}\, e^{\frac{N_0 B}{p_{k,n}}} \, E_{A_k}\left(\frac{N_0 B}{p_{k,n}}\right)\right) - \left(\lambda \, + \,  q_{i-1} \, T_f \, \varrho \right) \, \phi_{k,n} \, p_{k,n} \right] \nonumber
\end{eqnarray}
\end{figure*}

%%%%%%%%%%%%%%%%%%%%%%%%%%%%%%%%%%%% Equation at the top of the next page (37)
\begin{figure*}[!t]
\normalsize
\centering
\begin{equation}\label{eq:newform}
 \argmax_{p_{k,n},{\phi}_{k,n}} \left[\dfrac{\phi_{k,n}(\nu_k + 1)}{\theta_k}
  \, \ln\left(\frac{N_0 B}{p_{k,n}}\, e^{\frac{N_0 B}{p_{k,n}}} \, E_{A_k}\left(\frac{N_0 B}{p_{k,n}}\right)\right) - \left(\lambda \, + \,  q_{i-1} \, T_f \, \varrho \right) \, \phi_{k,n} \, p_{k,n} \right] 
\end{equation}
\vspace*{-5mm}
\hrulefill
\end{figure*}

Since Problem \eqref{eq:newform} is in a standard concave form, the Karush-Kuhn-Tucker (KKT) first order optimality conditions of\cite{Boyd04} may be used for finding the problem's optimal solution. The next two subsections deal with the updating process of the primal and dual variables.

%%%%%%%%%%%%%%%%%%%%%%%%%%%%%%%%%%%%%%%%%%%%%%%%%%%%%%%%%%%%
\subsubsection{Updating the Power and Subcarrier Allocation}
%%%%%%%%%%%%%%%%%%%%%%%%%%%%%%%%%%%%%%%%%%%%%%%%%%%%%%%%%%%%
For a fixed $\lambda$, $\nu_k$ and $q_{i-1}$, we may solve $\max_{{\bf P},{\pmb \phi}} \mathcal{L}({\bf P},{\pmb \phi},{\bm \nu},\lambda)$ in order to obtain the optimal power and subcarrier allocation. Therefore, the following condition is both necessary and sufficient for the power allocation's optimality:
\begin{equation}\label{eq:pcond}
 \left. \dfrac{\partial \, \mathcal{L}({\bf P},{\pmb \phi},{\bm \nu},\lambda)}{\partial p_{k,n}} \right|_{p_{k,n} = p_{k,n}^*} = 0,
\end{equation}
which is equivalent to finding the specific point given by \eqref{eq:derivativep}. This point can be computed using Newton's method.
%%%%%%%%%%%%%%%%%%%%%%%%%%%%%%%%%%%% Equation at the top of the next page (39)
\begin{figure*}[!t]
\normalsize
\centering
\begin{eqnarray}\label{eq:derivativep}
 - \dfrac{1}{p_{k,n}^2} \left( \dfrac{ \left(p_{k,n} + N_0 B\right) (\nu_k -1)}{\theta_k} + (\lambda \, + \, q_{i-1} \, T_f \, \varrho) p_{k,n}^2 - \left(\dfrac{N_0 B (\nu_k -1)}{\theta_k}\dfrac{E_{A_k -1}\left(\frac{N_0 B}{p_{k,n}}\right)}{E_{A_k}\left(\frac{N_0 B}{p_{k,n}}\right)}\right)  \right) = 0
\end{eqnarray}
\vspace*{-5mm}
\hrulefill
\end{figure*}

Once the optimal power allocation (${\bf P}^*$) has been calculated, the optimal subcarrier allocation may be obtained through:
\[
 \left. \dfrac{\partial \mathcal{L}({\bf P^*},{\pmb \phi},{\bm \nu},\lambda)}{\partial \phi_{k,n}} \right|_{\phi_{k,n} = \phi_{k,n}^*} = 0, 
\]
where we have 
\begin{eqnarray}\label{eq:subcarrieralloc_eee}
 \dfrac{\partial \mathcal{L}({\bf P^*},{\pmb \phi},{\bm \nu},\lambda)}{\partial \phi_{k,n}} &=& \left[\dfrac{(\nu_k+1)}{\theta_k} \ln\left(\mathcal{I}^{\texttt{RFC}}_{k,n}\right)\right. \\
 &&\left. - (\lambda - q_{i-1} \, T_f \, \varrho) \, p_{k,n} \dfrac{}{} \right] \nonumber \\
 &=& \Phi_{k,n} \left\{ \begin{array}{ll} 
                         < 0 & \text{ if } \phi_{k,n}^* = 0, \\
                         = 0 & \text{ if } \phi_{k,n}^* \in (0,1), \\
                         > 0 & \text{ if } \phi_{k,n}^* = 1.
                        \end{array} \nonumber
 \right.
\end{eqnarray}
Note that in \eqref{eq:subcarrieralloc_eee} the derivative is independent of $\pmb \phi$. Therefore, its value means that either the optimal value occurs at the boundaries of the feasible region, and thus $\mathcal{L}({\bf P}, {\pmb \phi}, {\bm \nu}, \lambda)$ must be a decreasing function within the feasible region, or the derivative is null and hence the optimal subcarrier allocation is obtained inside the feasible region. Since only a single user is allowed to transmit on each subcarrier, the following condition 
\begin{eqnarray}\label{eq:subcarrieralloc_cond}
 \phi_{k,n} = \left\{ \begin{array}{ll}
                       1, & \text{ if } \Phi_{k,n} = \max({\bm \Phi}_n) \\
                       0, & \text{ otherwise }
                      \end{array}
\right.
\end{eqnarray}
may be applied in a Gauss-Seidel fashion\cite{van_Loan_2012} when designing the iterative algorithm, where ${\bm \Phi}_n$ is the $n$th column of $\bm \Phi$.
Indeed, in the Gauss-Seidel-type iterative algorithms only a single dimension is considered at each iteration. Hence, this type of iterative algorithms are said to be sequential. For example, the iterative algorithm designed herein applies the condition \eqref{eq:subcarrieralloc_cond} to each subcarrier sequentially, rather than in parallel. This process is illustrated by the loop starting from Line 7 in Algorithm \ref{alg:DinkLagrange}, where the power allocation procedure is executed for each subcarrier of every user in the system and then the condition \eqref{eq:subcarrieralloc_cond} is applied to this particular subcarrier.

It is worth noting that since $\phi_{k,n}$ only assumes binary values and the condition \eqref{eq:subcarrieralloc_cond} implies that only a single user is assigned to each subcarrier. As a consequence, the constraints C3 and C5 are implicitly satisfied. Furthermore, the condition C4 is satisfied by the assumption that $\mathcal{U}_p({\pmb\phi},{\bf P})$ is a positive affine function, as shown in Lemma \ref{lemma:2}. Thus, the conditions C3-C5 may be omitted in the Lagrangian function \eqref{eq:lagrangian}.

%%%%%%%%%%%%%%%%%%%%%%%%%%%%%%%%%%%%%%%%%%%%%%%%%%%%%%%%%%%%
\subsubsection{Updating the Dual Variables}
%%%%%%%%%%%%%%%%%%%%%%%%%%%%%%%%%%%%%%%%%%%%%%%%%%%%%%%%%%%%
In order to update the dual variables $\lambda$ and ${\bm \nu}$, one may use the subgradient algorithm, whose equations are presented in \eqref{eq:lambda} and \eqref{eq:nu}. The parameters $\alpha_\lambda$ and $\alpha_\nu$ are the appropriate step sizes of the subgradient algorithm.

%%%%%%%%%%%%%%%%%%%%%%%%%%%%%%%%%%%% Equation at the top of the next page (41)
\begin{figure*}[!t]
\normalsize
\centering
\begin{eqnarray}\label{eq:lambda}
 \lambda(i+1) &=& \lambda(i) + \alpha_\lambda \dfrac{\partial \mathcal{L}({\bf P},{\pmb \phi},{\bm \nu},\lambda)}{\partial \, \lambda} \nonumber \\
  &=& \lambda(i) + \alpha_\lambda \left(P_{\max} - \sum\limits_{k=1}^K \sum\limits_{n=1}^N \phi_{k,n} p_{k,n}\right)
\end{eqnarray}
\vspace*{-5mm}
\hrulefill
\end{figure*}

%%%%%%%%%%%%%%%%%%%%%%%%%%%%%%%%%%%% Equation at the top of the next page (42)
\begin{figure*}[!t]
\normalsize
\centering
\begin{eqnarray}\label{eq:nu}
 \nu_k(i+1) &=& \nu_k(i) + \alpha_\nu \dfrac{\partial \mathcal{L}({\bf P},{\pmb \phi},{\bm \nu},\lambda)}{\partial \, \nu_k} \nonumber \\
 &=& \nu_k(i) + \alpha_\nu \left(\mathcal{C}_e^{k,\min} - \sum\limits_{n=1}^N -\dfrac{\phi_{k,n}}{\theta_k} \ln\left(\frac{N_0 B}{p_{k,n}}\, e^{\frac{N_0 B}{p_{k,n}}} \, E_{A_k}\left(\frac{N_0 B}{p_{k,n}}\right)\right)\right)
\end{eqnarray}
\vspace*{-5mm}
\hrulefill
\end{figure*}

\subsection{Dinkelbach-Lagrange Dual Decomposition Algorithm}
The algorithm developed for our EEE-optimal design of OFDMA networks under statistical delay-QoS provision is summarized in this section. The main loop of the proposed algorithm is composed by Dinkelbach's algorithm illustrated in Algorithm \ref{alg:Dink}. The Lagrange dual decomposition procedure is used for solving the inner loop, which is equivalent to solving Problem \eqref{eq:dink_solv}. The pseudo-code in Algorithm \ref{alg:DinkLagrange} implements the entire power and subcarrier allocation policies for our EEE maximization problem. The variables used throughout each algorithm are presented at the end of Algorithm \ref{alg:DinkLagrange}, while the identifiers in round brackets indicate to which procedure the variable belongs: (L) for the Lagrange dual decomposition method and (D) for Dinkelbach's method.

\begin{algorithm}[!h]
\small
\caption{Dinkelbach-Lagrange Dual Decomposition Algorithm for EEE-Optimal Design}\label{alg:DinkLagrange}
{ \linespread{1}
\noindent {\bf Input:} ${\bf P}$, ${\pmb \phi}$, $I_{dd}$, $\epsilon_{\lambda}$, $\epsilon_{\nu}$, $\epsilon_{Dink}$, $I_{Dink}$ \\
\noindent {\bf Output:} ${\bf{P}^*}$, ${\pmb \phi}^*$\\
begin \\
1.\hspace*{0.1in} Initialize ${\bf P}$, ${\pmb \phi}$; \\
2.\hspace*{0.1in} $i \leftarrow 0$; \\
3.\hspace*{0.1in} \underline{\textbf{while}} $i \leq I_{Dink}$ \textbf{or} $|F(q_i)| \leq \epsilon_{Dink}$ \\
4.\hspace*{0.25in} $q_i \leftarrow \eta_E^{\texttt{RFC}}$; \\
5.\hspace*{0.25in} $j \leftarrow 0$; \\
6.\hspace*{0.25in} \textbf{while} $j \leq I_{dd}$ \textbf{or} \big($\left| \lambda(j+1) - \lambda(i) \right| > \epsilon_{\lambda}$ \textbf{and} \\
\hspace*{0.78in} $|\min({\bm \nu}(j+1) - {\bm \nu}(j))| > \epsilon_{\nu}$\big) \\
7.\hspace*{0.40in} \textbf{for} $n$ \textbf{from} $1$ \textbf{to} $N$  \\
8.\hspace*{0.55in} \textbf{for} $k$ \textbf{from} $1$ \textbf{to} $K$  \\
9.\hspace*{0.70in} Find $p_{k,n}^*$ that satisfies \eqref{eq:pcond} or \eqref{eq:derivativep};   \\
10.\hspace*{0.49in} \textbf{end for} \\
11.\hspace*{0.49in} Obtain the optimal subcarrier allocation using  \\
\hspace*{0.65in}  \eqref{eq:subcarrieralloc_eee} and \eqref{eq:subcarrieralloc_cond}; \\
12.\hspace*{0.35in} \textbf{end for} \\
13.\hspace*{0.35in} Update the dual variables using \eqref{eq:lambda} and \eqref{eq:nu}; \\ 
14.\hspace*{0.35in} $j \leftarrow j +1$; \\ 
15.\hspace*{0.21in} \textbf{end while} \\
16.\hspace*{0.21in} $i \leftarrow i+1$; \\
17.\hspace*{0.07in} \textbf{end \underline{while}} \\
\hspace*{0in}---------------------------------------- \\
\footnotesize
\hspace*{0.0in} ${\bf P}$ = initial power allocation matrix;\\
\hspace*{0.0in} ${\bf P}^*$ = optimal power allocation;\\
\hspace*{0.0in} ${\pmb \phi}$ = initial subcarrier allocation matrix;\\
\hspace*{0.0in} ${\pmb \phi}^*$ = optimal subcarrier allocation;\\
\hspace*{0.0in} $I_{dd}$ = maximum number of iterations (L);\\
\hspace*{0.0in} $I_{Dink}$ = maximum number of iterations (D);\\
\hspace*{0.0in} $\epsilon_{Dink}$ = Dinkelbach algorithm precision (D);\\
\hspace*{0.0in} $\epsilon_{\lambda}$ = power allocation precision (L);\\
\hspace*{0.0in} $\epsilon_{\nu}$ = subcarrier allocation precision (L).}
%\hspace*{0.10in} $\mathbb{E}[\cdot]$ is the mathematical expectation.
%
\end{algorithm}

The underlined \textbf{while} in Line 3 represents the main Dinkelbach loop, while the non-underlined \textbf{while} in Line 6 corresponds to the Lagrange dual decomposition method's main loop or, alternatively to Dinkelbach's method's inner loop.

\section{Simulations and Numerical Results} \label{sec:numerical}
In order to illustrate the algorithm's performance in solving our EEE-maximization problem, numerical simulations were conducted. The  adopted simulation parameter values for the downlink of the OFDMA system considered are presented in Table~\ref{tab:ofdma_param}.

Aiming for comparing different scenarios associated with different solutions and for evaluating the impact of the parameter values on the solution of the EEE-maximization problem, in Table \ref{tab:scenarios} we summarize four different scenarios: the first two scenarios are simple and were used for investigating the impact of each system parameter on the result of the EEE-optimization problem. The third and fourth scenarios are more realistic, with a larger number of subcarriers, wider subcarrier bandwidth and more users. Therefore, they are more complex to deal with.

\begin{table}[htpb]
    \centering
    \caption{OFDMA system parameters} \label{tab:ofdma_param}
    \vspace{-.03mm}
	\small
    \begin{tabular}{ll}
        \hline
        \textbf{Parameters} & \textbf{Values}\\
        \hline
        \hline
        \multicolumn{2}{c}{\textit{OFDMA System (Downlink)}} \\
        \hline
        Total bandwidth              & $\mathcal{B} \in \{0.1;20\}$ [MHz] \\
        Number of subcarriers	     & $N = \{2; 3; 16; 32; 128\}$ \\
        Subcarrier bandwidth         & $B=\mathcal{B}/N$ [MHz] \\
        Frame duration               & $T_f=667\mu$s\\
        Noise  variance              & $N_0=10^{-12}$ $\left[\frac{\text{Watts}}{\text{Hertz}}\right]$\\
        Max. transmit power  & $P_{\max} = 30$ [dBm]\\
        Circuit power               & $P_\textsc{c} =\{20; 50\}$ [dBm]  \cite{Loodaricheh2014}\\
        PA inefficiency   & $\varrho = \frac{\textsc{papr}}{\xi} = 2.5$ \cite{Fettweis10} \\
	Minimum EC  & $C_{\textsc{e},{\texttt{RFC}}}^{k,\min} = 1$ $\left[\frac{\text{bits}}{T_f}\right]$\\
        \hline
        \multicolumn{2}{c}{\textit{Dinkelbach's Method's Parameters}} \\
        \hline
        Max. number of iterations & $I_{Dink} = 100$\\
        Expected precision & $\epsilon_{Dink} = 10^{-6}$\\
        \hline
        \multicolumn{2}{c}{\textit{Lagrange Dual Decomposition Parameters}} \\
        \hline
        Max. number of iterations & $I_{dd} = 500$\\
        Expected precisions & $\epsilon_{\lambda} = 10^{-6}$\\
                            & $\epsilon_{\nu} = 0.5$\\
        \hline
    \end{tabular}
\end{table}

\begin{table}[htpb]
    \centering
    \caption{Scenario Parameters} \label{tab:scenarios}
    \vspace{-.03mm}
		\small
    \begin{tabular}{ll}
        \hline
        \textbf{Parameters} & \textbf{Values}\\
        \hline
        \hline
        \multicolumn{2}{c}{\textit{Scenario 1}} \\
        \hline
        Number of users              & $K = 2$  \\
        Number of subcarriers              & $N=3$ \\
        QoS-exponent vector              & ${\bm \theta} = [\theta_1 =0.1, \theta_2 = 0.25]$\\
        Total bandwidth           & $\mathcal{B} = 0.1$ [MHz] \\
        Circuit power               & $P_\textsc{c} =20$ [dBm]  \\
        \hline
        \multicolumn{2}{c}{\textit{Scenario 2}} \\
        \hline
        Number of users              & $K = 2$  \\
        Number of subcarriers              & $N=2$ \\
        QoS-exponent vector              & ${\bm \theta} = [\theta_1 = 0.1, \theta_2 = 0.25]$\\
        Total bandwidth           & $\mathcal{B} = 0.1$ [MHz] \\
        Circuit power               & $P_\textsc{c} =20$ [dBm]  \\
        \hline
        \multicolumn{2}{c}{\textit{Scenario 3}} \\
        \hline
        Number of users              & $K = [2, 10, 20]$  \\
        Number of subcarriers              & $N=[3, 16, 32]$ \\
        QoS-exponent              & $\theta_k \sim \mathcal{U}(0,1),\forall k$ \\
        Total bandwidth           & $\mathcal{B} = 20$ [MHz] \\
        Circuit power               & $P_\textsc{c} =20$ [dBm] \\
	\hline
        \multicolumn{2}{c}{\textit{Scenario 4}} \\
        \hline
        Number of users              & $K = 50$  \\
        Number of subcarriers              & $N=128$ \\
        QoS-exponent              & $\theta_k \sim \mathcal{U}(0,1),\forall k$ \\
        Total bandwidth           & $\mathcal{B} = 20$ [MHz] \\
        Circuit power               & $P_\textsc{c} =50$ [dBm]  \cite{Loodaricheh2014}\\
        \hline
    \end{tabular}
\end{table}

\begin{figure}[htpb]
 \centering
 \subfigure[EEE contour]{
 \label{fig:EEE_contour}
 \includegraphics[width=\linewidth]{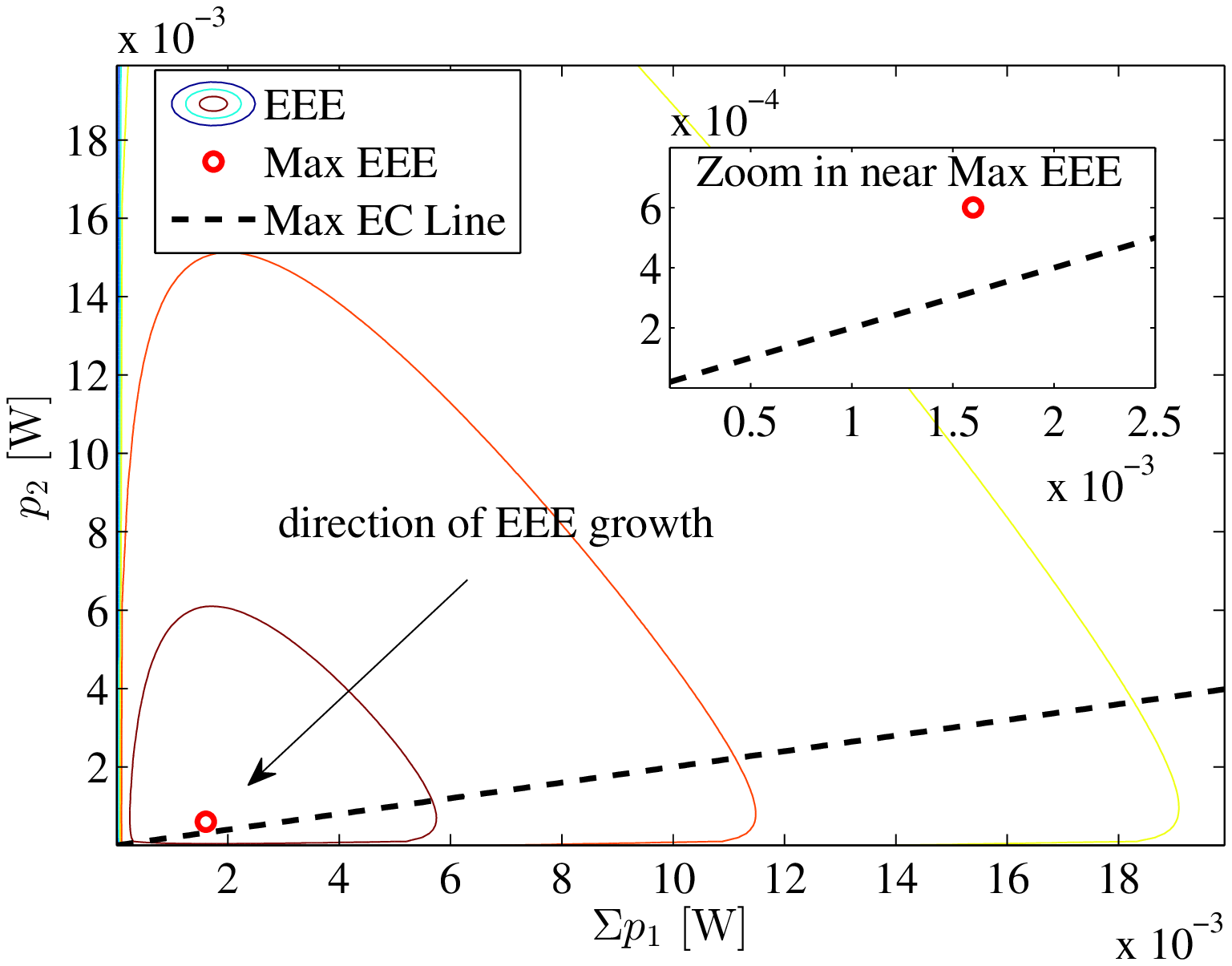}
 }
 \subfigure[EEE surface]{
 \label{fig:EEE_surface} 
 \includegraphics[width=\linewidth]{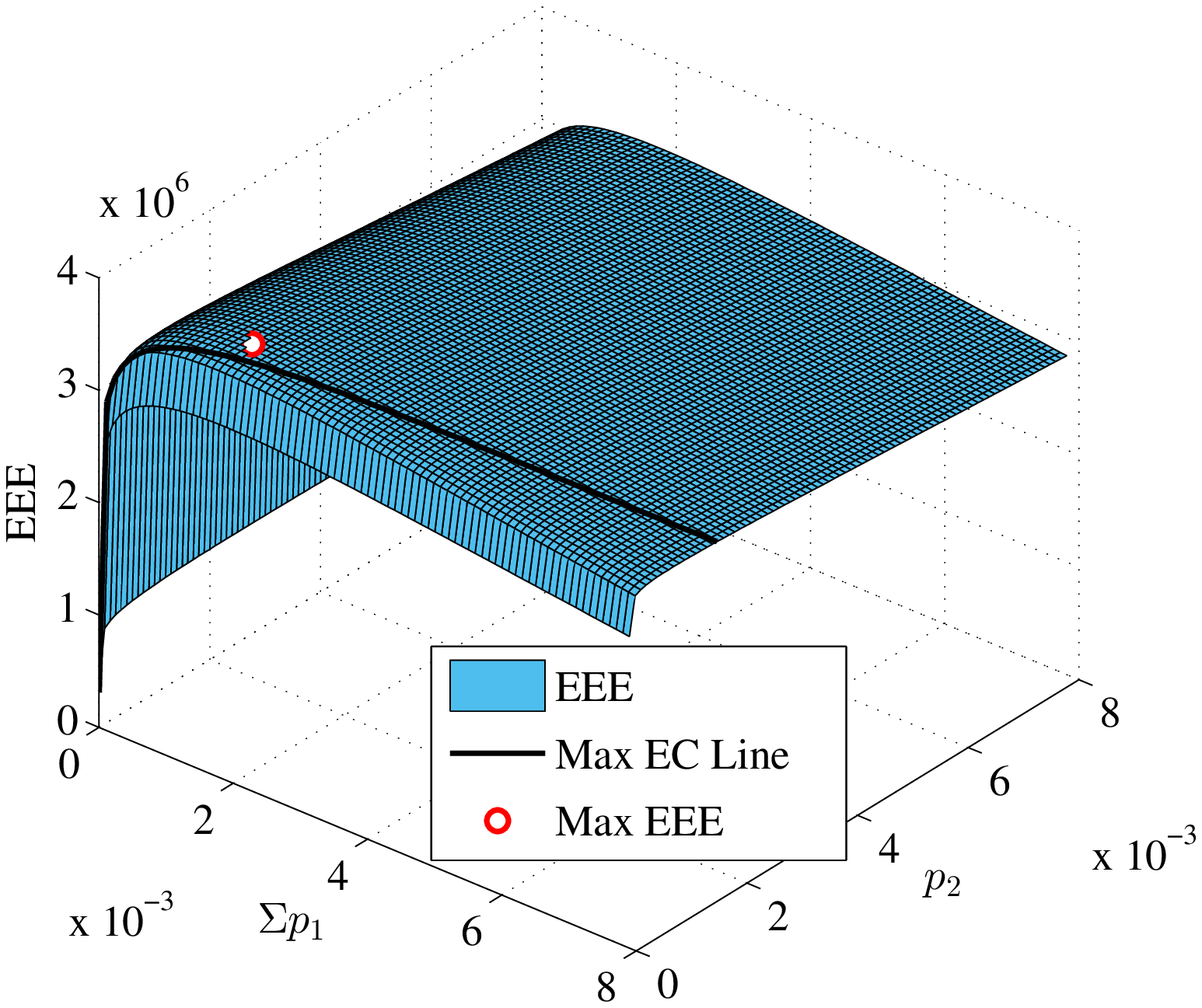}
 }
 \vspace{-1mm}
 \caption{EEE contour (a) and surface (b) for Scenario 1 of Table \ref{tab:scenarios}. The black dashed line in (a) or the black solid line in (b) represent the optimal power allocation policy that achieves the maximum effective capacity. The red circle shows the optimal power allocation that achieves the maximum EEE. Note that $\sum p_1$ represents the total power that is used for transmission from the BS to User 1 over all the subcarriers allocated to this user.}\label{fig:eee_scenario1}
\end{figure}

In order to observe the relationship between the EEE and the EC, we present Fig. \ref{fig:eee_scenario1} which illustrates the contour plot of the EEE surface with respect to the total transmission power of User 1 and 2 in Scenario 1 of Table \ref{tab:scenarios}. Since we have $\theta_1 < \theta_2$, two subcarriers are allocated to User 1, while User 2 only has a single subcarrier to transmit information. The figure also presents the maximum EC line (black dashed line). For a given $P_{\max}$ the black dashed line shows the optimal power allocation policy that achieves the maximum EC.
It is noteworthy that the maximum EC line is not far from the maximum EEE point, as seen in the zoomed-in part of Fig. \ref{fig:eee_scenario1}.

\begin{figure}[tbp]
 \centering
 \subfigure[]{
 \label{fig:convergence_alg2_in_transmit_power_scenario_1}
 \includegraphics[width=\linewidth]{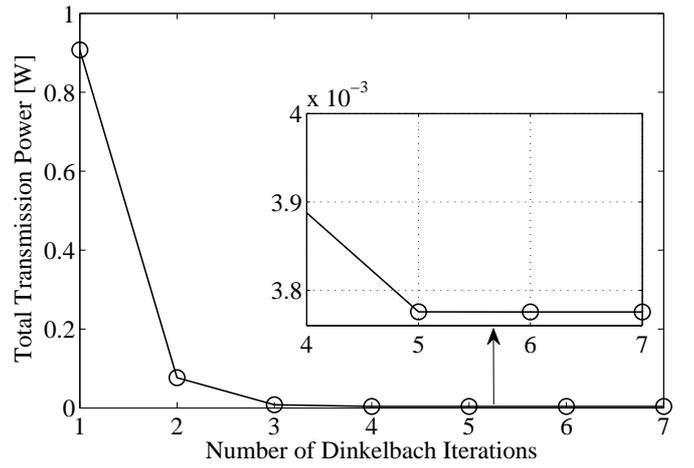}
 }
 \subfigure[]{
 \label{fig:convergence_alg2_in_EEE_scenario_1} 
 \includegraphics[width=\linewidth]{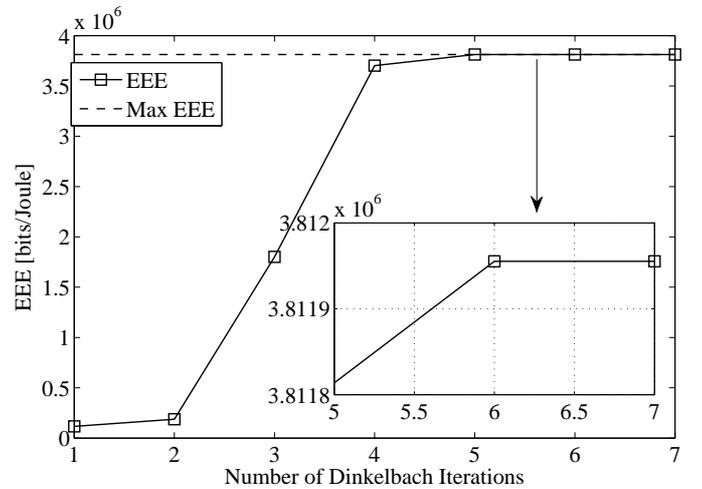}
 } 
 \caption{Typical convergence profile of Algorithm 2 in terms of the total transmission power (a) and EEE (b) for Scenario 1 of Table \ref{tab:scenarios}. }\label{fig:eee_1}
\end{figure}
\begin{figure}[tbp]
 \centering
 \subfigure[]{
 \label{fig:convergence_alg2_in_transmit_power_scenario_4}
 \includegraphics[width=\linewidth]{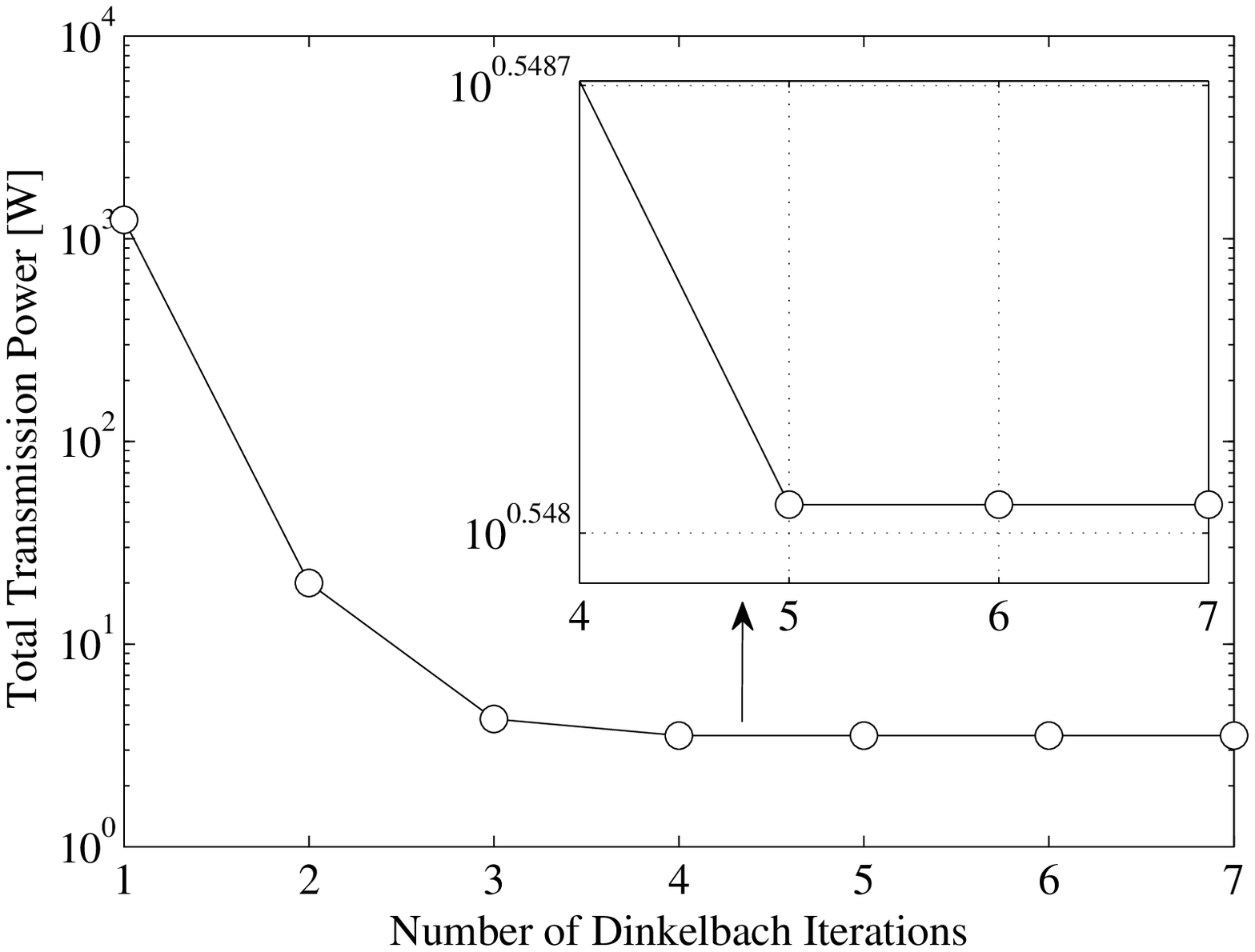}
 }
 \subfigure[]{
 \label{fig:convergence_alg2_in_EEE_scenario_4}
 \includegraphics[width=\linewidth]{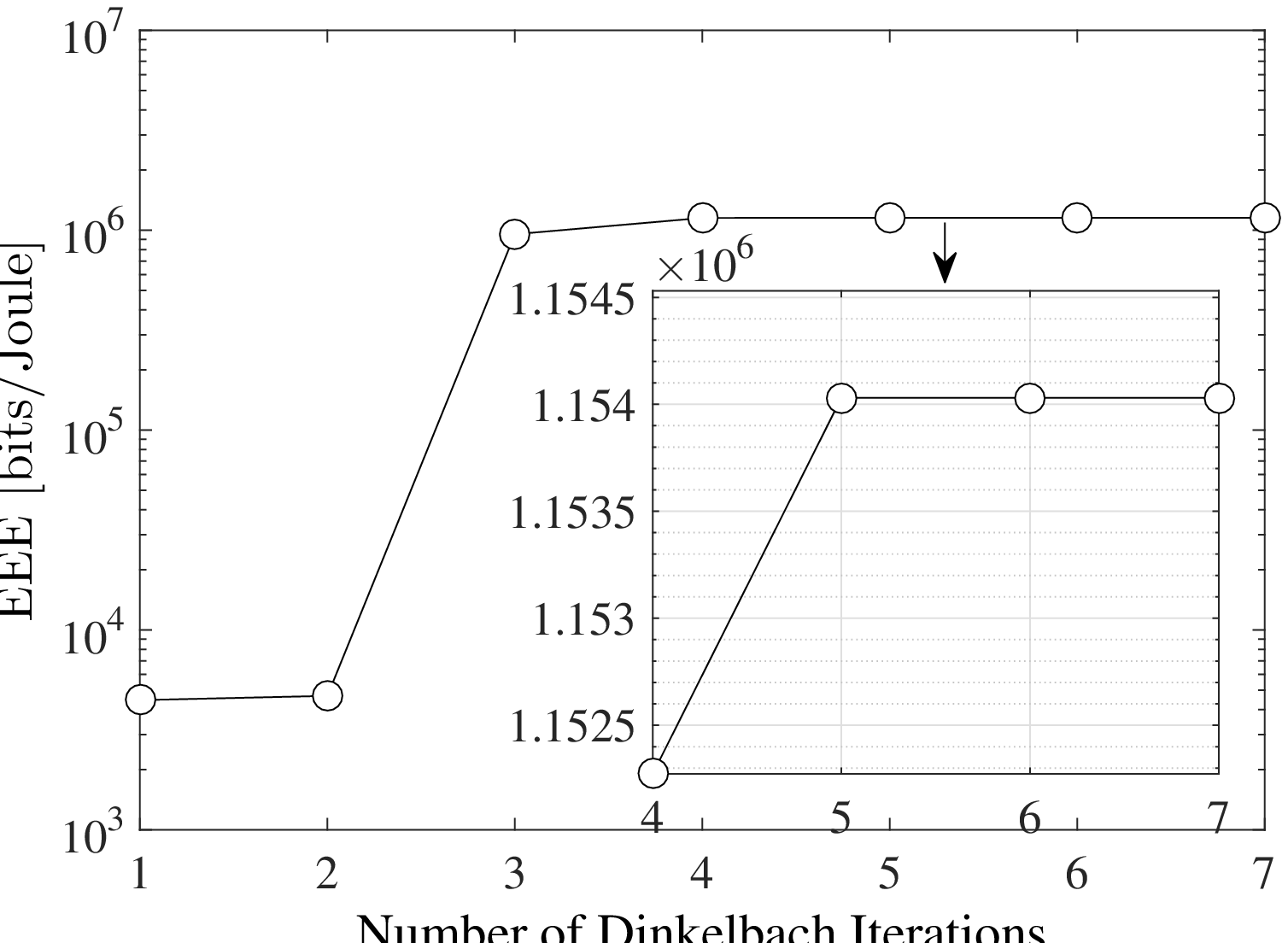}
 }
 \caption{Typical convergence profile of Algorithm \ref{alg:DinkLagrange} in terms of total transmission power (a) and achievable EEE (b) for a realistic system configuration of Scenario 4 in Table \ref{tab:scenarios}.}\label{fig:128}
\end{figure}

Fig. \ref{fig:eee_1} shows the convergence of Algorithm 2 in terms of the total transmission power [Fig. \ref{fig:convergence_alg2_in_transmit_power_scenario_1}] and EEE [Fig. \ref{fig:convergence_alg2_in_EEE_scenario_1}] for Scenario 1 of Table \ref{tab:scenarios}. Note that the maximum EEE curve in Fig.~\ref{fig:eee_1} was found through an exhaustive search method considering both the subcarrier allocation and power allocation domains. We observe from Fig. \ref{fig:eee_1} that the algorithm requires only 6 iterations to converge\footnote{The convergence behavior of the Scenario 2 and Scenario 3 of Table  \ref{tab:scenarios} is omitted here, since it is similar to that of Scenario 1.}. 
Moreover, Fig. \ref{fig:128} depicts the typical convergence profile for the proposed Algorithm \ref{alg:DinkLagrange} in terms of the total transmission power and EEE for a more realistic system configuration, namely for Scenario 4 of Table \ref{tab:scenarios}, with a product of $K N = 6400$. The maximum achievable EEE is not shown in Fig.~\ref{fig:128}, since there are $2^N$ possible subcarrier allocation matrices and hence an exhaustive search becomes computationally prohibitive. We can see that full convergence to the EEE-optimal design is achieved by the proposed algorithm after 5 Dinkelbach iterations within a precision of $\epsilon = 10^{-6}$. Additionally, it is noteworthy that the EEE values achieved for this realistic scenario are significantly lower than those of the less realistic Scenario 1. For example, by comparing the achievable EEE shown in Fig. \ref{fig:eee_1} (under Scenario 1, with product $KN = 6$) and Fig. \ref{fig:128} (under the realistic Scenario 4), we can see that the EEE of Scenario 1 is almost 100 times higher than that of Scenario 4. This difference is mainly due to the different circuitry power consumption values $P_\textsc{c}$, which has been increased from $20 \textrm{dBm} = 100 \textrm{mW}$ in Scenario 1 to $50 \textrm{dBm} = 100 \textrm{W}$ in Scenario 4. This result corroborates our previous discussions concerning \eqref{eq:2terms_pc} on the importance of  the fixed circuitry power consumption $P_\textsc{c}$ at BSs.
%
%Moreover, Fig.~ \ref{fig:eee_3} shows similar results considering the more complex Scenario 3. Observe from Fig.~\ref{fig:eee_3} that the algorithm takes 8 iterations to achieve its convergence in this context, which indicates an increase of around $30\%$ compared to Scenario~1. The maximum achievable EEE is not shown in Fig.~\ref{fig:eee_3}, since there are $2^N$ possible subcarrier allocation matrices and hence an exhaustive search becomes computationally prohibitive.
%\begin{figure}[htpb]
% \centering 
% \subfigure[]{
% \label{fig:convergence_alg2_in_transmit_power_scenario_3}
% \includegraphics[width=.48\textwidth]{figs/fig_3a.eps}
% }
% \subfigure[]{
% \label{fig:convergence_alg2_in_EEE_scenario_3}
% \includegraphics[width=.48\textwidth]{figs/fig_3b.eps} 
% }
% \caption{Typical convergence profile of Algorithm 2 in terms of the total transmission power (a) and EEE (b) for Scenario 3 of Table \ref{tab:scenarios}. }\label{fig:eee_3}
%\end{figure}

Furthermore, by jointly considering multiple representative simulation scenarios, it is possible to evaluate the average number of iterations required by the proposed algorithm for achieving convergence. To elaborate a little further, Table \ref{tab:res} presents the average number of iterations (over 100 realizations) at which the proposed algorithm converges (i.e., $\epsilon \leq 10^{-6}$) under different values of $(K, N)$. It is noteworthy that the increase in problem dimensions, represented by the product $KN$ of Scenario 3, only imposes a modest impact on the number of  Dinkelbach iterations required. As seen from Algorithm 1 and (22), the computational complexity per Dinkelbach iteration is roughly the same in terms of complexity order, with $K$ and $N$ only slightly affecting the number of simple summations. Hence, the computational complexity required by Algorithm \ref{alg:DinkLagrange} to achieve convergence also increases modestly with $KN$.
\begin{table}
 \centering
 \caption{Average number of Dinkelbach iterations to achieve convergence under different number of users and subcarriers of Scenario 3}\label{tab:res}
 \begin{tabular}{cc|c}
 \hline
  \bf \# of Users & \bf \# of Subcarriers & \bf \# of Iterations \\
  \hline 
  $K=2$ & $N=3$ & $5.97$ \\
  $K=10$ & $N=16$ & $6.86$ \\
  $K=20$ & $N=32$ & $7.41$\\
 \hline
 \end{tabular}
\end{table}

\begin{figure}[t]
\centering \includegraphics[width=\linewidth]{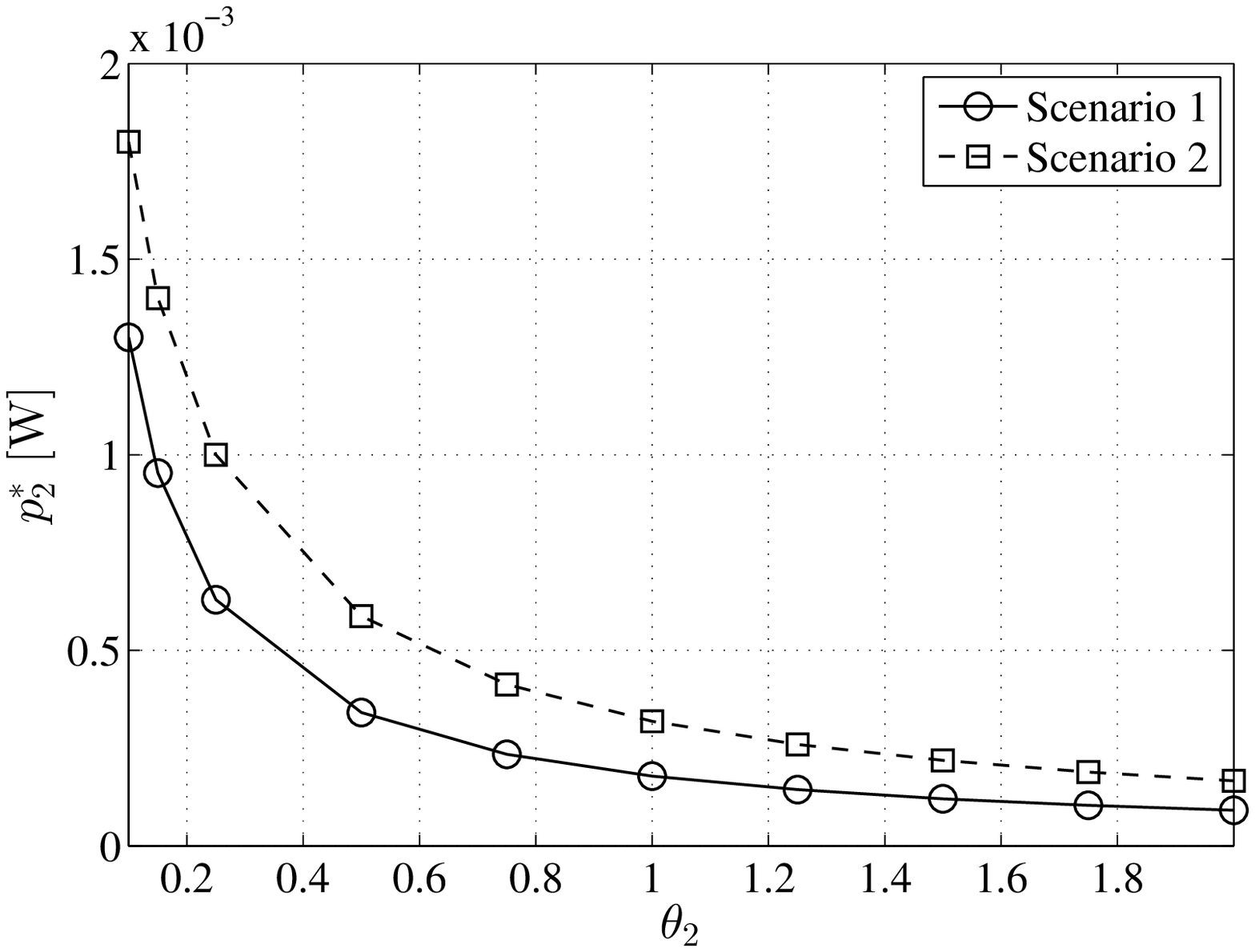}
\caption{Optimal power allocation for User 2 considering different values of $\theta_2$ within the interval $[0.1; 2]$, while $\theta_1$ is fixed to $0.1$, and the values of the other parameters are the same as those of Scenario 1 and Scenario 2.}
\label{fig:thetaopt}
\end{figure}
In order to gain further insights into the EEE-maximization problem considered, the impact of three parameters of paramount importance  are evaluated by considering the associated optimal transmit power of User 2, i.e., $p_2^*$. The first parameter examined was the QoS-exponent $\theta$, which has a direct relationship both to the maximum delay bound $d_{\text{max}}$ and to the probability of not exceeding this bound, $\varepsilon$. Hence, in Fig. \ref{fig:thetaopt}  we show the optimal power allocation for User 2 considering different values of its QoS-exponent $\theta_2$ in the interval of $[0.1; 2]$, while keeping the QoS-exponent of User 1 at $\theta_1=0.1$. The values of the other parameters are the same as those of both Scenario 1 and Scenario 2. Note that for Scenario 1, which has three subcarriers, two of them are allocated to User 1, while User 2 receives information through only one of the OFDMA subcarriers. The results of Fig. \ref{fig:thetaopt} demonstrate the impact of different values of $\theta$ on the optimal power allocation that achieves the maximum EEE. Since $\theta$ is related to both the maximum delay bound $d_{\text{max}}$ and its violation probability $\varepsilon$, physically it can characterize both stringent delay-QoS requirement (larger $\theta$) as well as loose delay-QoS requirement (smaller $\theta$). For instance, $\theta = 1$ can represent $\varepsilon = 10\%$ probability of violating a delay-limit of $d_{\max} = 2.3$ seconds, or $\varepsilon = 50\%$ probability of violating $d_{\max}= 0.69$ seconds; $\theta = 0.23$ can indicate $\varepsilon = 10\%$ probability of violating $d_{\max} = 10$ seconds, as predicted by \eqref{eq:theta_bound}. The following conclusion can be drawn directly from Fig. \ref{fig:thetaopt}: the optimal power allocation policy has an exponential decay dependence with respect to the QoS-exponent $\theta$, implying that a lower delay tolerance, i.e, a smaller $d_{\text{max}}$ or a larger $\theta$ in \eqref{eq:theta_bound}, requires a lower transmit power to achieve the optimal EEE and vice-versa.

\begin{figure}[t]
 \centering
 \includegraphics[width=\linewidth]{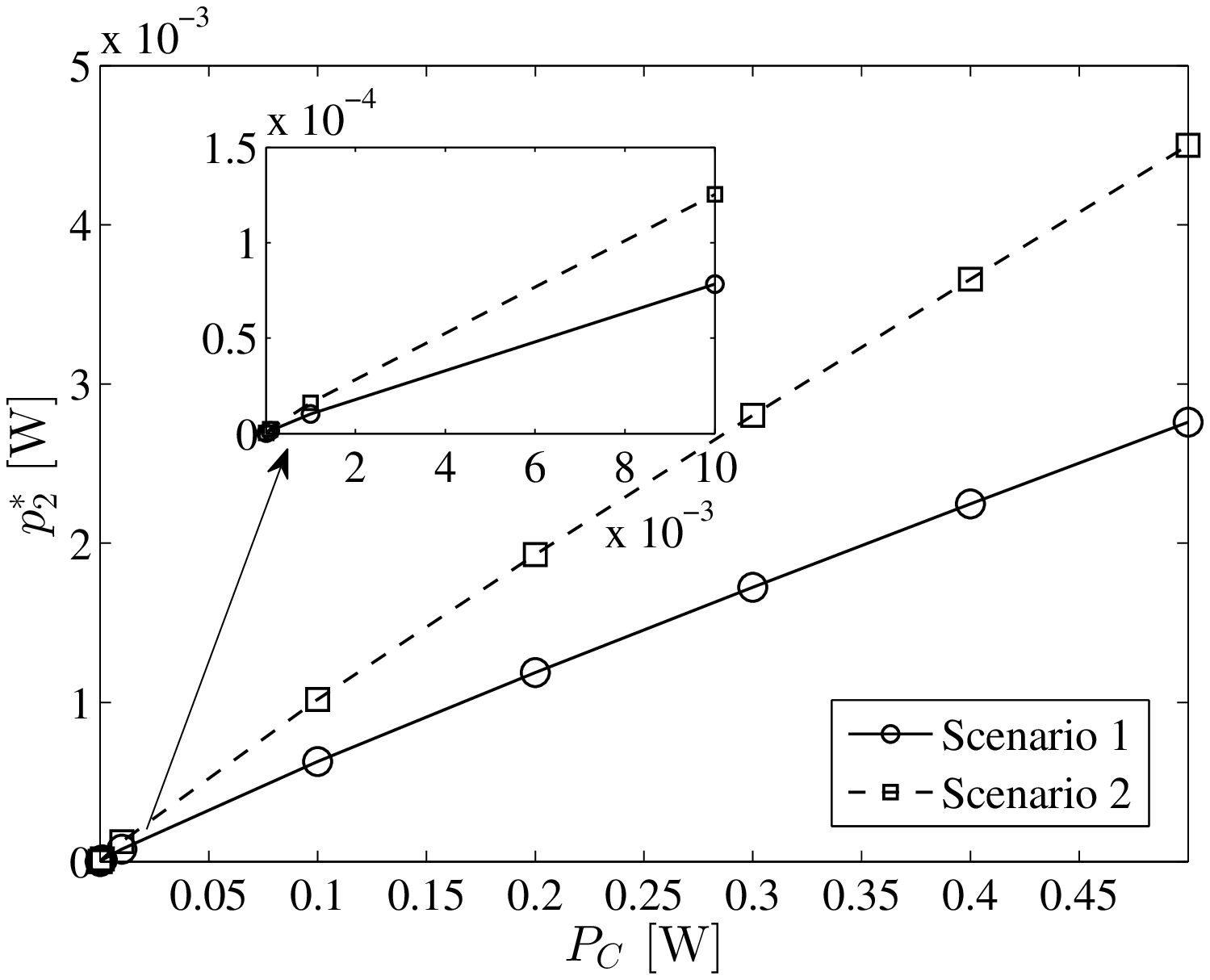} 
 \caption{Optimal power allocation for User 2 considering different values of $P_\textsc{c}$ in both Scenario 1 and Scenario 2.}\label{fig:pcopt}
\end{figure}
The second parameter studied is the circuitry power $P_\textsc{c}$. Fig.~\ref{fig:pcopt} depicts the optimal power allocation value for User 2 in both Scenario 1 and Scenario 2, where the EEE is maximized. In contrast to the impact of the QoS-exponent, as the circuitry power consumption increases, the optimal power allocation value increases linearly with it.

\begin{figure}[t]
 \centering \includegraphics[width=\linewidth]{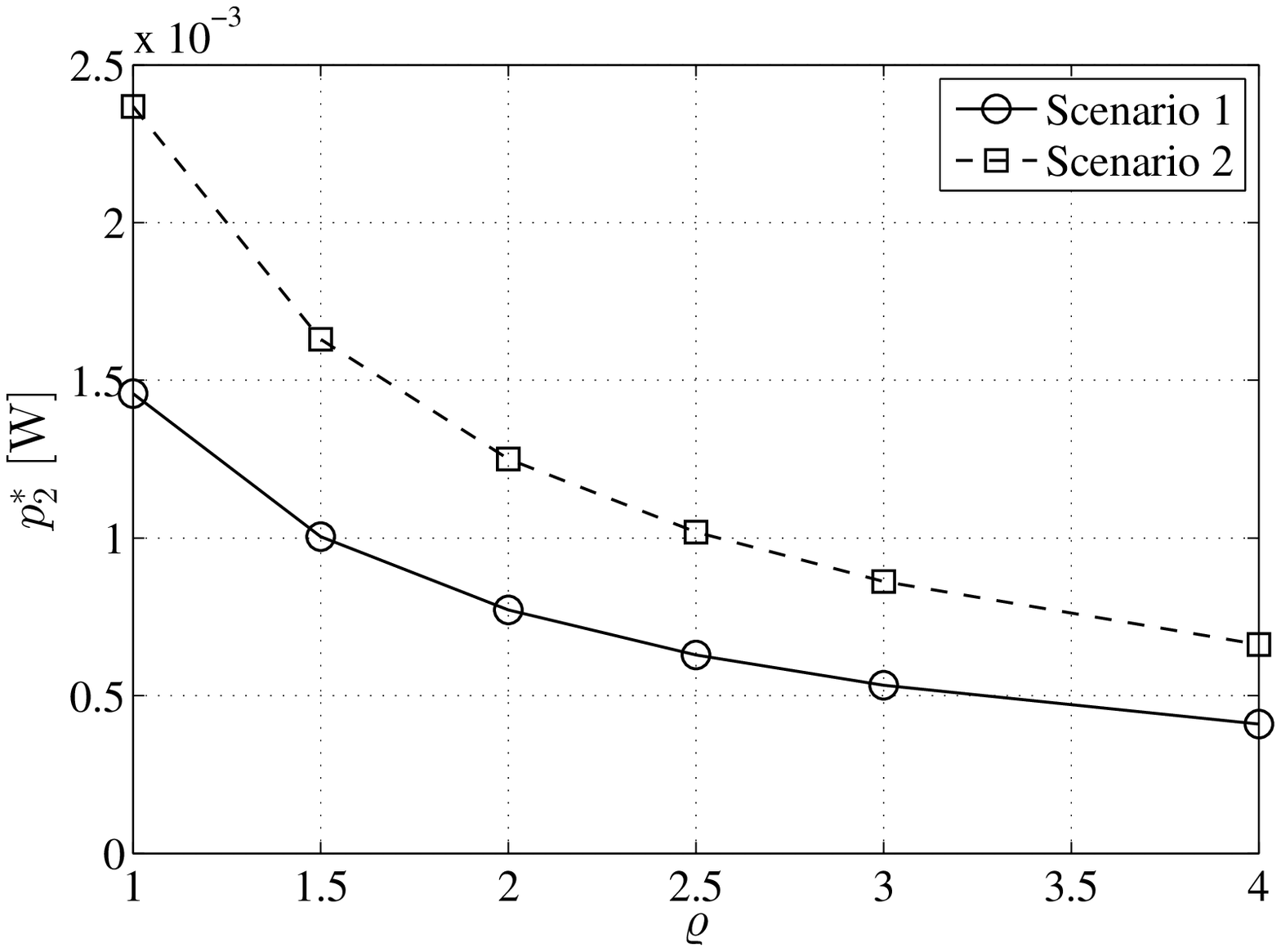} 
 \caption{Optimal power allocation for User 2 considering different values of $\varrho$ in both Scenario 1 and Scenario 2.}\label{fig:varrhoopt}
\end{figure}

The third parameter investigated is the PA inefficiency $\varrho$. From its definition given in Section \ref{ofdma_power_consumption_model}, we know that $\varrho$ is directly proportional to the PAPR value and inversely proportional to the drain efficiency of the PA. Fig. \ref{fig:varrhoopt} shows the optimal power allocation policy for User 2 in both Scenario 1 and Scenario 2. As we may observe in this figure, the optimal power allocation value decays exponentially with the PA inefficiency, but smoother than the trend is for the QoS-exponent.
In fact, if we consider the scenario of a single user and a single subcarrier, the EEE function will result in a well-known bell-shaped curve. Hence, increasing either the PA inefficiency or the QoS-exponent basically shifts the optimum point to the left of the original optimum point, while increasing the circuitry power shifts the optimum point to the right of the original one.

% \begin{table}[htpb]
%     \centering
%     \caption{Scenario 4 Systems Parameters} \label{tab:sc4}
%     \vspace{-.03mm}
% 	\small\colb{
%     \begin{tabular}{ll}
%         \hline
%         \textbf{Parameters} & \textbf{Values}\\
%         \hline
%         \hline
%         \multicolumn{2}{c}{\textit{OFDMA System (Downlink)}} \\
%         \hline
%         Total bandwidth              & $\mathcal{B} =20$ [MHz] \\
%         Number of users              & $K = 50$  \\
%         Number of subcarriers              & $N=128$ \\
%         Subcarrier bandwidth         & $B=156.25$ [KHz] \\
%         Frame duration               & $T=667\mu$s\\
%         Noise  variance              & $N_0=10^{-12}$ $\left[\frac{\text{Watts}}{\text{Hertz}}\right]$\\
%         Max. transmit power  & $P_{\max} = 30$ [dBm]\\
%         Circuit power               & $P_\textsc{c} = 100$ [W]  \cite{Loodaricheh2014}\\
%         PA inefficiency   & $\varrho = \frac{\textsc{papr}}{\xi} = 2.5$ \\
%         QoS exponent              & $\theta_k = \mathcal{U}(0,1),\forall k$ \\
%         Minimum effective capacity  & $C_{\textsc{e},{\texttt{RFC}}}^{k,\min} = 1$ $\left[\frac{\text{bits}}{\text{frame}}\right]$\\
% 	\hline
%     \end{tabular}}
% \end{table}

%%%%%%%%%%%%%%%%%%%%%%%%%%%%%%%%%%%%%%%%%%%%%%%%%%%%%%%%%%%%%%%%%%%%%%%%
\section{Conclusions}\label{sec:conclusions}
%%%%%%%%%%%%%%%%%%%%%%%%%%%%%%%%%%%%%%%%%%%%%%%%%%%%%%%%%%%%%%%%%%%%%%%%
In this paper we have demonstrated the concavity of the EC function and the quasi-concavity of the EEE function. The relaxed EEE-maximization problem was reformulated for using Dinkelbach's method, which is capable of solving a more tractable parameterized version of the original fractional programming problem. The Lagrange dual decomposition method was invoked to solve the sub-optimization-problem that emerges in the inner loop of Dinkelbach's method.
Our numerical simulation results have demonstrated that the proposed algorithm is capable of converging to the optimal solution in a small number of iterations, even under realistic scenarios associated with large system dimensions quantified in terms of the product of the number of users and subcarriers, i.e. $KN$. We also offered an investigation concerning the system parameters in order to quantify how each of the three key parameters impacts the EEE function maximization, which facilitates a deeper understanding of the importance of these parameters in circuitry and infrastructure design.

%========================================================
\appendices
\section{Proof of Lemma \ref{lemma:1}}\label{app:1}
%========================================================
\begin{proof}
We commence the proof by showing that: 
 \begin{eqnarray*}
\mathcal{C}(\underline{p}_{k,n}, \underline{\phi}_{k,n}) =  -\frac{\underline{\phi}_{k,n}}{\theta_k}  \, \ln\left( \frac{N_0 \underline{B}_{k,n}}{\underline{p}_{k,n}} e^{\frac{N_0 \underline{B}_{k,n}}{\underline{p}_{k,n}}} E_{A_k} \left[ \frac{N_0 \underline{B}_{k,n}}{\underline{p}_{k,n}}\right] \right)
 \end{eqnarray*}
is concave. Since $\mathcal{C}(\cdot)$ is twice differentiable, the second-order test may be applied to verify its concavity. Thus, the Hessian matrix ${\bf H}$ of $\mathcal{C}(\cdot)$ is:
\begin{eqnarray*}
 {\bf H}_\mathcal{C} &=& \left[ { \begin{array}{cc} 
  \dfrac{\partial^2 \, \mathcal{C}\left(p_{k,n}, \phi_{k,n}\right)}{\partial p_{k,n}^2} & \dfrac{\partial^2 \, \mathcal{C}\left(p_{k,n}, \phi_{k,n}\right)}{\partial p_{k,n} \partial \phi_{k,n}} \\
 \dfrac{\partial^2 \, \mathcal{C}\left(p_{k,n}, \phi_{k,n}\right)}{\partial \phi_{k,n} \partial p_{k,n}} & \dfrac{\partial^2 \, \mathcal{C}\left(p_{k,n}, \phi_{k,n}\right)}{\partial \phi_{k,n}^2}
 \end{array} }\right] \\
 &=& \left[ {\begin{array}{cc} \textsc{c}_{pp} & \textsc{c}_{p\phi} \\ \textsc{c}_{\phi p} & \textsc{c}_{\phi\phi} \end{array}} \right],
\end{eqnarray*}
where $\textsc{c}_{pp}$, $\textsc{c}_{p\phi}$, $\textsc{c}_{\phi p}$ and $\textsc{c}_{\phi\phi}$ are defined in \eqref{eq:cpp}.
\begin{figure*}[t]
 \normalsize
 \begin{eqnarray}
 \begin{aligned}\label{eq:cpp}
 \textsc{c}_{pp} &= \dfrac{\phi_{k,n} \left(e^{-\frac{2 N_0 B}{p_{k,n}}} p_{k,n} - e^{-\frac{N_0 B}{p_{k,n}}} (N_0 B + (2 - A_k) p_{k,n}) \, E_{A_k}\left[\frac{N_0 B}{p_{k,n}}\right] - (2 N_0 B + A_k \, p_{k,n}) E_{A_k}\left[\frac{N_0 B}{p_{k,n}}\right]^2 \right) }{\theta_k p_{k,n}^3 E_{A_k}\left[\frac{N_0 B}{p_{k,n}}\right]^2}, \\
 \textsc{c}_{p\phi} &= \textsc{c}_{\phi p} = \dfrac{(N_0 B + p_{k,n}) \, E_{A_k}\left[\frac{N_0 B}{p_{k,n}}\right] - (N_0 B) \,\, E_{A_k -1}\left[\frac{N_0 B}{p_{k,n}}\right] }{\theta_k p_{k,n}^2 E_{A_k}\left[\frac{N_0 B}{p_{k,n}}\right]},  \\
 \textsc{c}_{\phi\phi} &= 0.   
 \end{aligned}
 \end{eqnarray}
 \vspace*{4pt}
 \hrulefill
\end{figure*}
From \eqref{eq:cpp} we may conclude that ${\bf H}$ is a $(2 \times 2)$-element symmetric matrix and the following statements are equivalent \cite[Theorem 1.10, p. 11]{berman2003}:

\begin{enumerate}
 \item  ${\bf H}$ is semidefinite negative;
 \item All principal minors of ${\bf H}$ are nonpositive.
\end{enumerate}
In fact it may be easily verified that both principal minors $M_i$ of ${\bf H}$ are nonpositive for any $p_{k,n} \geq 0$ and $\phi_{k,n} \in [0,1]$:
\begin{eqnarray*}
 M_1 &=& \textsc{c}_{pp} \leq 0, \\
 M_2 &=& {-\textsc{c}_{p\phi} \textsc{c}_{\phi p}} =  -(\textsc{c}_{p\phi})^2.
\end{eqnarray*}
Naturally, $M_2$ is nonpositive since it is the negative counterpart of a quadratic term. However, it is not easy to observe that $\textsc{c}_{pp} \leq 0$ by simply checking the expression in \eqref{eq:cpp}. In order to show that the inequality holds, let us consider, without loss of generality, that $\phi_{k,n}=1$. Fig. \ref{fig:regionplot} illustrates the regions, specified by $N_0 B$, $p_{k,n}$ and $A_k$, where $M_1$ is satisfied. Other alternative methods for demonstrating the validity of this inequality include demonstrating that the second derivatives regarding $p_{k,n}$, $N_0 B$ and $A_k$ are negative. However, this is omitted here due to space limitations.
%\vspace{-5mm}

\begin{figure}[htpb]
 \centering
\includegraphics[width=.6\linewidth]{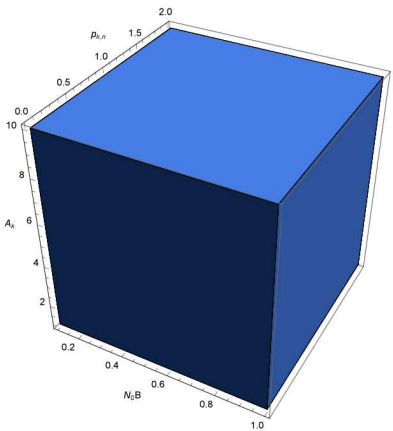}
\vspace{-4mm}
 \caption{Illustration of the 3D region plot, where the inequality $\textsc{c}_{pp} \leq 0$ holds, i.e. we have a polyhedric convex set. The polyhedron base is formed by $N_0 B$ and $p_{k,n}$ axes, while its height is $A_k$.}\label{fig:regionplot}
\end{figure}

As shown in Fig. \ref{fig:regionplot}, $M_1$ holds for any $p_{k,n} \geq 0$. Since $M_1$ grows linearly with $\phi_{k,n}$ and  $\phi_{k,n}$ is nonnegative, the only condition for $M_1$ to hold is $p_{k,n} \geq 0$. Therefore, the Hessian is semidefinite negative, which implies that $\mathcal{C}(\underline{p}_{k,n}, \underline{\phi}_{k,n})$ is concave.

Finally, according to \cite[p.79]{Boyd04} (operations that preserve convexity), the following statement is true: if $\mathcal{C}(\underline{p}_{k,n}, \underline{\phi}_{k,n})$  is concave, then $\mathcal{C}_e(\underline{\bf P},\underline{\bm \phi}, {\bm \theta})$ is concave, because it is the nonnegative weighted sum of concave functions.
\end{proof}

\begin{biography}[{\includegraphics[width=1in,height=1.25in,clip,keepaspectratio]{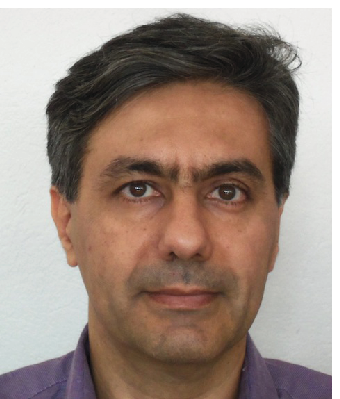}}]
{\bf Taufik Abr\~ao} (M'97-SM'12) (\url{http://www.uel.br/pessoal/taufik}) received the B.S., M.Sc., and Ph.D. degrees in electrical engineering from the Polytechnic School of the University of São Paulo, São Paulo, Brazil, in 1992, 1996, and 2001, respectively. Since March 1997, he has been with the Communications Group, Department of Electrical Engineering, Londrina State University, Paraná, Brazil, where he is currently an Associate Professor in Telecommunications and the Head of the Telecomm. \& Signal Processing Lab. In 2012, he was an Academic Visitor with the Southampton Wireless Research Group, University of Southampton, Southampton, U.K. From 2007 to 2008, he was a Post-doctoral Researcher with the Department of Signal Theory and Communications, Polytechnic University of Catalonia (TSC/UPC), Barcelona, Spain. He has participated in several projects funded by government agencies and industrial companies. He is involved in editorial board activities of six journals in the telecommunications area and has served as TPC member in several symposiums and conferences. He has also served as an Editor for IEEE Communications Surveys \& Tutorials since 2013 and IET Journal of Engineering since 2014. He is a member of SBrT and a senior member of IEEE. His current research interests include communications and signal processing, specially the multi-user detection and estimation, MC-CDMA and MIMO systems, cooperative communication and relaying, resource allocation, as well as heuristic and convex optimization aspects of 3G and 4G wireless systems. He has supervised 20 M.Sc. and  2 Ph.D. students, as well as 2 postdocs, co-authored nine book chapters on mobile radio communications and more than 170 research papers published in specialized/international journals and conferences.
\end{biography}

\begin{biography}[{\includegraphics[width=1in,height=1.25in,clip,keepaspectratio]{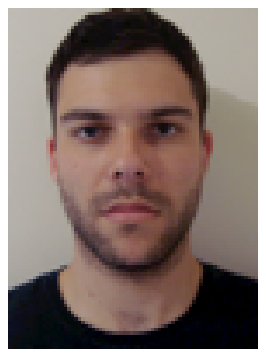}}]
{\bf Lucas Dias Hiera Sampaio} (S'10-M'16) received the B.S. and M.Sc. degrees in Computer Science from the State University of Londrina, Paraná, Brazil, in 2010 and 2011, respectively, and the Ph.D. degree in Electrical Engineering from the Polytechnic School of the University of São Paulo, Brazil in 2015. Since 2015 he holds a postdoc position at the Telecomm. \& Signal Processing Lab., Department of Electrical Engineering, Londrina State University, Brazil, and a temporary contract as assistant professor in Computer Science at the same university. His main research area includes telecommunications systems, wireless networks, cooperative networks, optimization and game theory applications in telecommunications, heuristics methods and green communications.
\end{biography}

\begin{IEEEbiography}[{\includegraphics[width=1in,height=1.25in,clip,keepaspectratio]{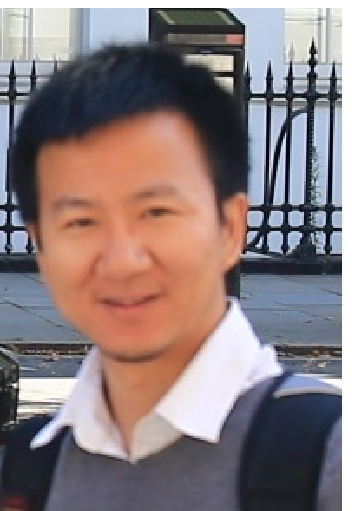}}] {Shaoshi Yang}
(S'09-M'13) received his B.Eng. degree in Information Engineering from Beijing University of Posts and Telecommunications (BUPT), Beijing, China in Jul. 2006, his first Ph.D. degree in Electronics and Electrical Engineering from University of Southampton, U.K. in Dec. 2013, and his second Ph.D. degree in Signal and Information Processing from BUPT in Mar. 2014. He is now working as a Postdoctoral Research Fellow in University of Southampton, U.K. From November 2008 to February 2009, he was an Intern Research Fellow with the Communications Technology Lab (CTL), Intel Labs, Beijing, China, where he focused on Channel Quality Indicator Channel (CQICH) design for mobile WiMAX (802.16m) standard. His research interests include MIMO signal processing, green radio, heterogeneous networks, cross-layer interference management, convex optimization and its applications. He has published in excess of 25 research papers on IEEE journals. 

Shaoshi has received a number of academic and research awards, including the prestigious Dean's Award for Early Career Research Excellence at University of Southampton, the PMC-Sierra Telecommunications Technology Paper Award at BUPT, the Electronics and Computer Science (ECS) Scholarship of University of Southampton, and the Best PhD Thesis Award of BUPT. He is a member of IEEE/IET, and a junior member of Isaac Newton Institute for Mathematical Sciences, Cambridge University, U.K. He also serves as a TPC member of several major IEEE conferences, including \textit{IEEE ICC, GLOBECOM, VTC, WCNC, PIMRC, ICCVE, HPCC}, and as a Guest Associate Editor of \textit{IEEE Journal on Selected Areas in Communications.} (\url{https://sites.google.com/site/shaoshiyang/}) 
\end{IEEEbiography}

\begin{IEEEbiography}[{\includegraphics[scale=0.315]{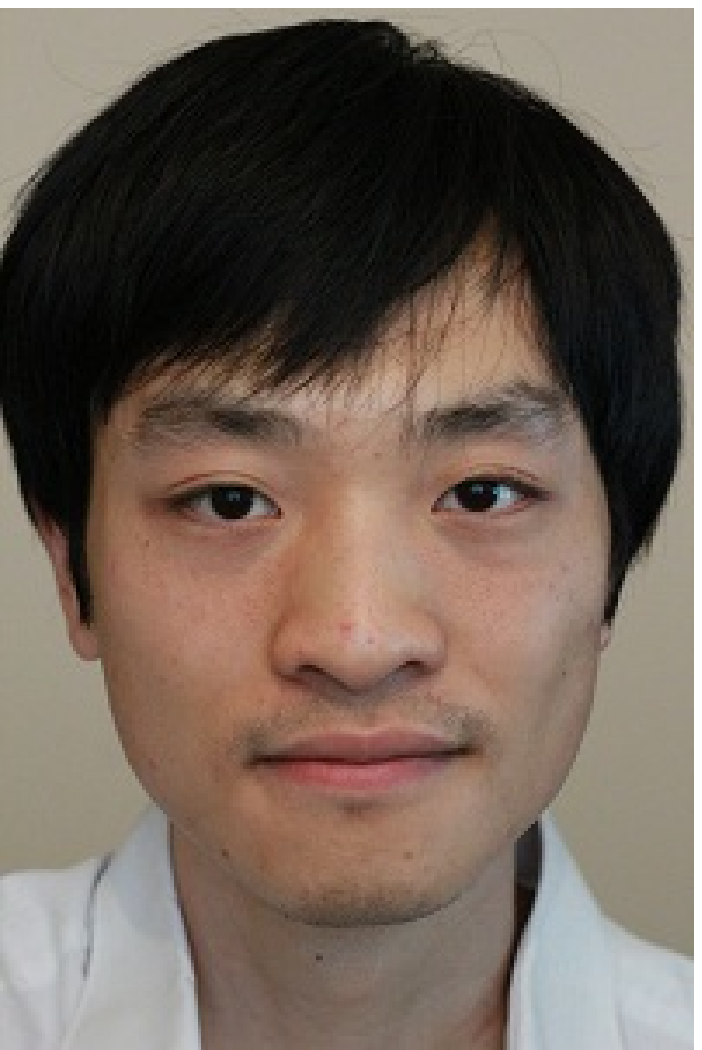}}]{Kent Tsz Kan Cheung}~(S'09) received his B.Eng. degree~(first-class honors) in Electronic Engineering and his Ph.D. degree in Electronics and Electrical Engineering, both from the Univeristy of Southampton, Southampton, U.K., in 2009 and 2015, respectively. He was a recipient of the EPSRC Industrial CASE award in 2009, and was involved with the Core 5 Green Radio project of the Virtual Centre of Excellence in Mobile and Personal Communications (Mobile VCE). His research interests include energy-efficiency, multi-carrier MIMO communications, cooperative communications, resource allocation and optimization.\end{IEEEbiography}

\begin{biography}[{\includegraphics[width=1in,height=1.25in,clip,keepaspectratio]{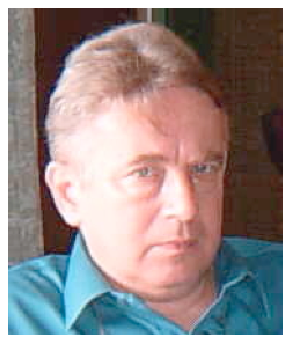}}]
{\bf Paul Jean Etienne Jeszensky} (SM'07) (\url{http://www.lcs.poli.usp.br/~pjj})
was born in Nancy-France. He received the B. S., M. S. and Ph. D. degrees, all in Electrical Engineering from EPUSP-Escola Politécnica da Universidade de São Paulo, in 1972, 1981, and 1989, respectively. From 1974 to 1978 he was with FDTE-Fundação para o Desenvolvimento Tecnológico da Engenharia as a member of Technical Staff on the development of Communication Systems. From 1978 to 1990 he taught courses part-time at EPUSP as an Instructor, and was also with FDTE and FUSP-Fundação de Apoio à Universidade de São Paulo, where he worked as a Technical Leader in various development programs of Communication Systems sponsored by the industry and the government. Since 1990 he has been with EPUSP, where he was a full-time Associate Professor since 1992 and a Full Professor since 2005. 

He was the Co-ordinator of LCS-Laboratório de Comunicações e Sinais of EPUSP in 1995-1997, 1999-2001 and 2003-2005, the General Co-Chairmen of the International Telecommunications Symposium (ITS'98), the General Chairman of the IEEE 9th International Symposium on Spread Spectrum Techniques and Applications (ISSSTA'06), and the General Co-ordinator for the cooperation agreements between Escola Politécnica and  Ericsson do Brasil, Motorola do Brasil and Telesp-Celular, since 1999. He was a visiting professor at UPC-Universitat Politécnica de Catalunya, Barcelona-Spain in 1995 and at TUB-Technical University of Budapest, Hungary in 2001. He is an author of more than 140 research papers published in periodicals and symposiums. He also published the book Sistemas Telefônicos (Editora Manole, 2004).
\end{biography}

\begin{IEEEbiography}[{\includegraphics[width=1in,height=1.25in,clip,keepaspectratio]{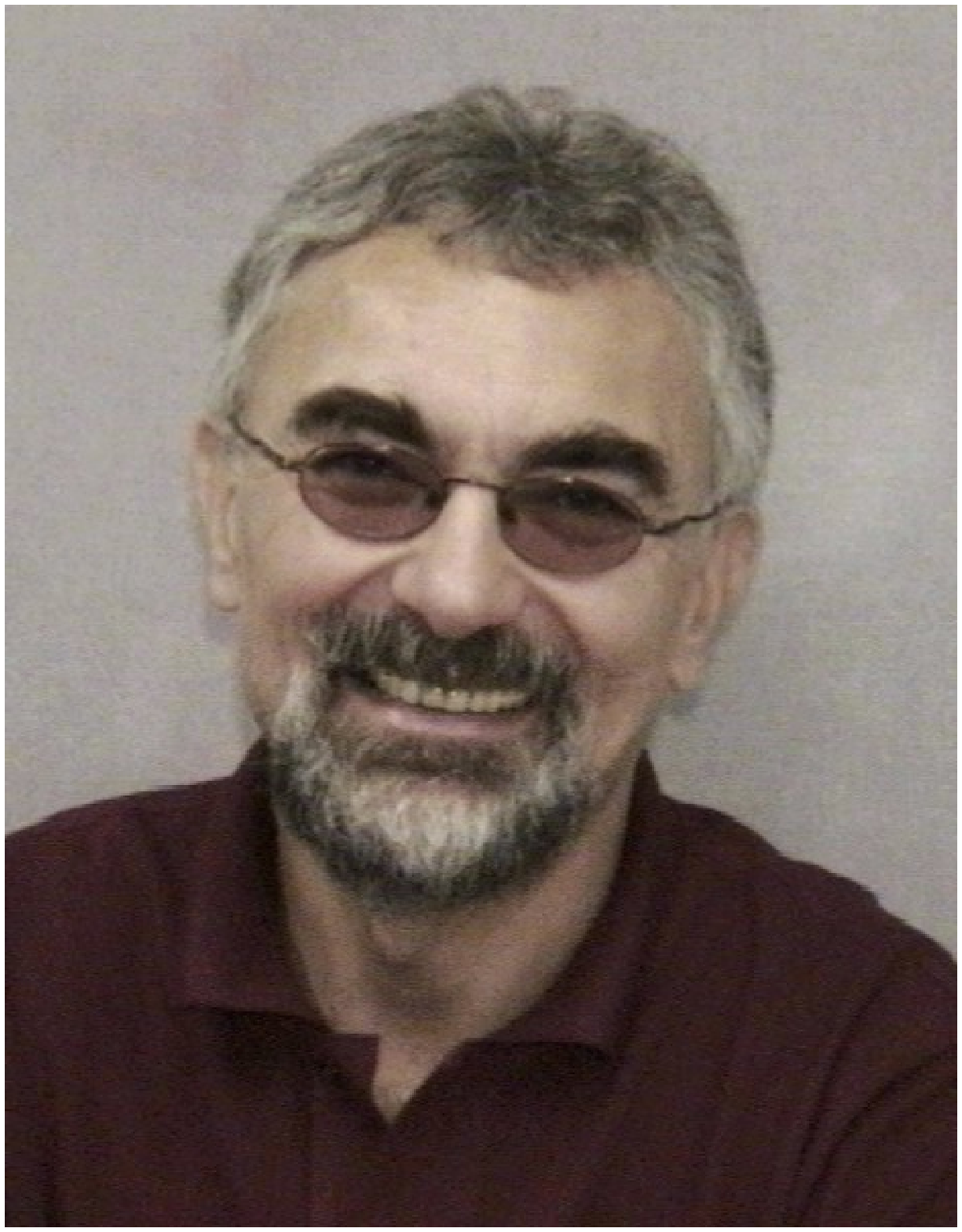}}] {Lajos Hanzo}
(M'91-SM'92-F'04) received his degree in electronics in
1976 and his doctorate in 1983. He was awarded an honorary
doctorate by the Technical University of
Budapest in 2009 and by the University of Edinburgh in 2015.  During his 39-year career in telecommunications he has held
various research and academic posts in Hungary, Germany and the
UK. Since 1986 he has been with the School of Electronics and Computer
Science, University of Southampton, UK, where he holds the Chair in
Telecommunications.  He has successfully supervised about 100 PhD students,
co-authored 20 John Wiley/IEEE Press books on mobile radio
communications totalling in excess of 10 000 pages, published 1500+
research entries at IEEE Xplore, acted both as TPC and General Chair
of IEEE conferences, presented keynote lectures and has been awarded a
number of distinctions. Currently he is directing a 60-strong
academic research team, working on a range of research projects in the
field of wireless multimedia communications sponsored by industry, the
Engineering and Physical Sciences Research Council (EPSRC) UK, the
European Research Council's Advanced Fellow Grant and the Royal
Society's Wolfson Research Merit Award.  He is an enthusiastic
supporter of industrial and academic liaison and he offers a range of
industrial courses. 

He is also a Fellow of the Royal Academy of Engineering, of the Institution
of Engineering and Technology (IET), and of the European Association for Signal
Processing (EURASIP). He is a Governor of the IEEE VTS.  During
2008 - 2012 he was the Editor-in-Chief of the IEEE Press and a Chaired
Professor also at Tsinghua University, Beijing. He 
has 24 000+ citations. For further information on research in progress and associated
publications please refer to \url{http://www-mobile.ecs.soton.ac.uk} 
\end{IEEEbiography}

\end{document}